\documentclass[amsmath, amssymb, aps, prl, reprint, floatfix, superscriptaddress]{revtex4-2}

\usepackage[T1]{fontenc}
\usepackage[utf8]{inputenc}
\usepackage{graphicx}
\usepackage{xcolor}
\definecolor{linkscolor}{RGB}{10,55,130}
\usepackage{bm}
\usepackage{braket}
\usepackage{times}
\usepackage{physics}
\usepackage{units}
\usepackage{upgreek}
\usepackage[percent]{overpic}
\usepackage{CJK}
\usepackage[pdftex, colorlinks=true, linkcolor=linkscolor, citecolor=linkscolor, urlcolor=linkscolor, breaklinks=true]{hyperref}

\begin{document}
\begin{CJK*}{UTF8}{bsmi}

\title{Floquet Engineering of a Quasiequilibrium Superradiant Phase Transition in Landau Polaritons}

\author{Wen-Hua Wu (吳文華)}
\affiliation{Applied Physics Graduate Program, Smalley-Curl Institute, Rice University, Houston, Texas 77005, USA}
\affiliation{Department of Electrical and Computer Engineering, Rice University, Houston, Texas 77005, USA}

\author{Fuyang Tay}
\affiliation{Department of Physics, Columbia University, New York, NY 10027, USA}
\affiliation{Department of Chemistry, Columbia University, New York, NY 10027, USA}

\author{Mengqian Che}
\affiliation{Department of Electrical and Computer Engineering, Rice University, Houston, Texas 77005, USA}

\author{Andrey Baydin}
\affiliation{Department of Electrical and Computer Engineering, Rice University, Houston, Texas 77005, USA}
\affiliation{Smalley-Curl Institute, Rice University, Houston, Texas 77005, USA}
\affiliation{Rice Advanced Materials Institute, Rice University, Houston, Texas 77005, USA}

\author{Junichiro Kono}
\affiliation{Department of Electrical and Computer Engineering, Rice University, Houston, Texas 77005, USA}
\affiliation{Smalley-Curl Institute, Rice University, Houston, Texas 77005, USA}
\affiliation{Rice Advanced Materials Institute, Rice University, Houston, Texas 77005, USA}
\affiliation{Department of Physics and Astronomy, Rice University, Houston, Texas 77005, USA}
\affiliation{Department of Materials Science and NanoEngineering, Rice University, Houston, Texas 77005, USA}

\author{David Hagenm\"{u}ller}
\email{david.hagenmuller@ipcms.unistra.fr}
\affiliation{IPCMS (UMR 7504), Universit\'{e} de Strasbourg and CNRS, Strasbourg, France}

\date{\today}

\begin{abstract}
Superradiant phase transitions (SRPTs), characterized by photon condensation and macroscopic matter polarization, are forbidden in equilibrium for homogeneous fields by no-go theorems. Here, we show that Floquet driving can circumvent this constraint in a Landau polariton system consisting of a two-dimensional electron gas coupled to a terahertz cavity in a DC magnetic field. An off-resonant AC magnetic field modulates the cyclotron frequency and light--matter coupling strength while leaving the diamagnetic term unchanged, generating an additional DC coupling contribution. This drives the system across a critical threshold into a superradiant phase, characterized by photon condensation and Landau-level polarization in the ground state of the Floquet Hamiltonian. This quasiequilibrium approach offers a route to SRPTs distinct from driven-dissipative schemes.
\end{abstract}

\maketitle
\end{CJK*}
Recent progress in cavity quantum electrodynamics has spurred interest in the interplay of collective many-body phenomena and strong light--matter interactions in solid-state platforms~\cite{Garcia-Vidal2021,Bloch2022,Schlawin2022,Hubener2024}. A paradigmatic example is provided by the Dicke model, which describes $N$ two-level systems with transition frequency $\omega_{0}$ coupled to a single cavity mode with strength $\Omega$~\cite{Dicke1954}. At the critical point $\Omega = 0.5,\omega_{0}$, where the lower-polariton mode softens to zero energy, the system undergoes a superradiant quantum phase transition (SRPT)  in the limit $N \to \infty$~\cite{HeppLieb1973,WangHioe1973,Carmichael1973}, characterized by macroscopic photon occupation and matter polarization in the ground state~\cite{EmaryPRL2003,EmaryBrandes2003,Kirton2019}.

In realistic systems, however, the minimal-coupling Hamiltonian includes a diamagnetic term quadratic in vector potential, $\boldsymbol{A}^{2}$, with strength $D$. For a spatially uniform $\boldsymbol{A}$, the Thomas--Reiche--Kuhn (TRK) sum rule enforces $D \ge \Omega^{2}/\omega_{0}$, leading to a no-go theorem that forbids photon condensation in equilibrium~\cite{Rzazewski1975,NatafCiuti2010}. This constraint extends to a broad class of models with homogeneous coupling~\cite{BialynickiBirulaRzazewski1979,GawedzkiRzazewski1981,HaynEmaryBrandes2012,Todorov2012,BambaOgawa2014,Tufarelli2015,RousseauFelbacq2017,Rokaj2018} and can be viewed as a manifestation of gauge invariance~\cite{KnightAharonovHsieh1978}. While additional dipole--dipole interactions in the Hamiltonian can produce ferroelectric~\cite{Keeling2007,DeBernardis2018_2,Stokes2020,PhysRevResearch.7.013271} or crystallization~\cite{Vukics2015} transitions, these occur without photon condensation.

Several strategies have been proposed to circumvent these no-go theorems, including inhomogeneous cavity vector potentials~\cite{NatafCiuti2010_2,NatafCiuti2011a,Bamba2016,NatafBasko2019,Andolina2020,Guerci2020,RomanRocheLuisZueco2021,Manzanares2022}, where the SRPT is accompanied by a static magnetic instability, multilevel emitters~\cite{Hayn2011,Baksic2013}, and Dicke analogs in magnetic compounds featuring magnon condensation~\cite{Li2018Observation,BambaLiMarquezPeracaKono2022,Liu2023,MarquezPeracaEtAl2024,KimEtAl2024}. Nevertheless, it has been repeatedly emphasized that reliable low-energy models require careful Hilbert space truncation~\cite{Viehmann2011,Abedinpour2011,Chirolli2012,DeBernardis2018,DiStefano2019,Andolina2019,Li2020,Garziano2020,Savasta2021,Dmytruk2021,RomanRoche2022,Li2022,Andolina2022NonPerturbative}.
To date, an equilibrium SRPT accompanied by photon condensation has not been observed. Such a transition is expected only under stringent conditions, notably requiring a large orbital susceptibility~\cite{NatafBasko2019,Andolina2020,Guerci2020,Manzanares2022}. 

Instead, recent efforts have gone beyond equilibrium by incorporating drive and dissipation~\cite{Ritsch2013RMP,DallaTorre2013PRA}. Driven-dissipative Dicke simulators were theoretically proposed~\cite{Dimer2007} and realized in a series of experiments~\cite{Baumann2010,Baumann2011Dicke,Brennecke2013PNAS,Baden2014PRL}. Although often termed SRPTs, these transitions are fundamentally nonequilibrium; the light--matter coupling is engineered via cavity-mediated Raman processes, that inherently involve external pumping and photon loss~\cite{Kirton2019}. Their critical properties, including critical exponents~\cite{Nagy2010,Klinder2015PNAS}, finite-size scaling~\cite{Konya2012PRA}, and dephasing effects~\cite{KirtonKeeling2017PRL}, have been widely explored.

\begin{figure*}[htbp]
    \centering
    \begin{minipage}[t]{0.33\textwidth}
        \vspace{0pt}
        \includegraphics[width=\linewidth]{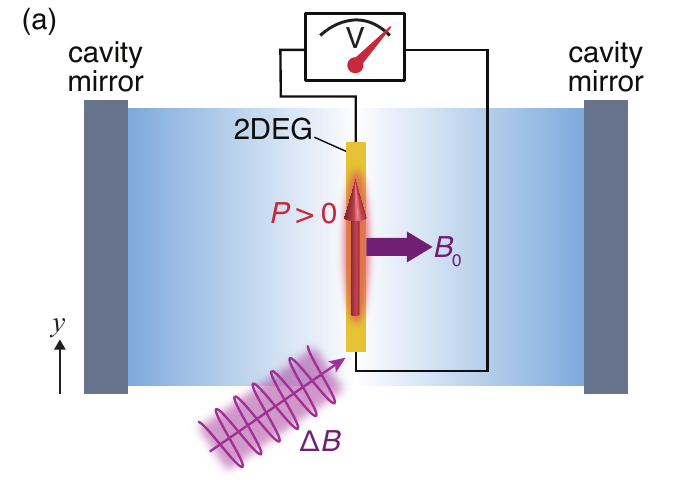}
    \end{minipage}\hfill 
    \begin{minipage}[t]{0.35\textwidth}
        \vspace{7pt}
        \includegraphics[width=\linewidth]{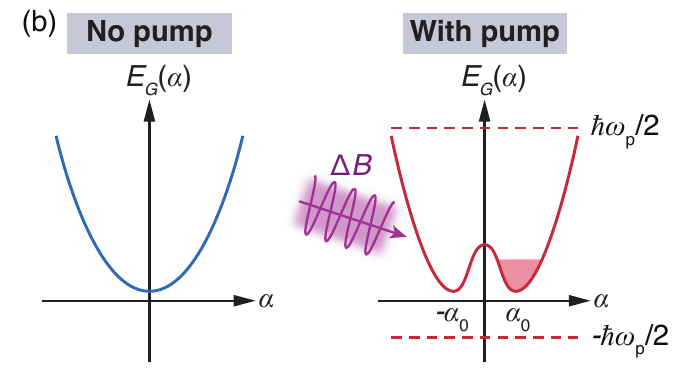}
    \end{minipage}\hfill 
    \begin{minipage}[t]{0.3\textwidth}
        \vspace{4pt}
        \includegraphics[width=\linewidth]{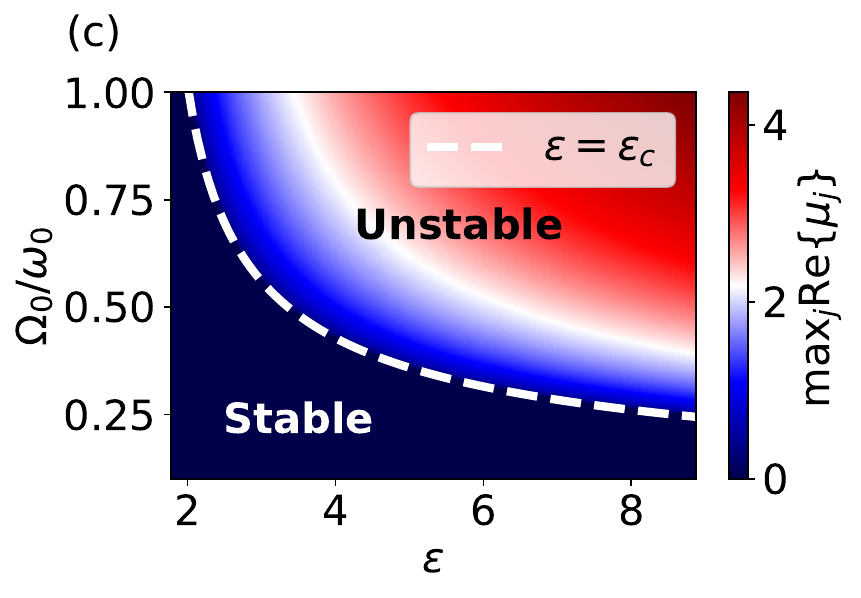}
    \end{minipage}
    \caption{\textbf{Floquet SRPT in a Landau polariton system.}
(a)~A two-dimensional electron gas in an optical cavity is subjected to a perpendicular magnetic field consisting of a static component $B_0$ and a periodic modulation $\Delta B$. In the superradiant phase, a macroscopic in-plane polarization $P$ emerges via spontaneous symmetry breaking.
(b)~Ground-state energy $E_\text{G}(\alpha)$ versus photonic field $\alpha$. Without driving, a single minimum at $\alpha=0$ (no SRPT, due to the TRK sum rule) appears. Under periodic modulation, two symmetric minima at $\pm \alpha_0$ emerge. Photon condensation selects one minimum via spontaneous parity breaking, signaling a quasiequilibrium SRPT. Dashed lines at $\pm \hbar \omega_\text{p}/2$ mark the Floquet Brillouin zone boundaries.
(c)~Maximum real part of the Floquet exponents, $\max_{j}\mathrm{Re}\{\mu_{j}\}$, versus the relative magnetic-field modulation amplitude $\varepsilon=\Delta B/B_0$ and equilibrium coupling strength $\Omega_0/\omega_0$. Positive values signal a Floquet instability. The white dashed line marks the analytical threshold, $\eta^2\varepsilon = 1+\omega_0^2/(4\Omega_0^2)$, separating stable and unstable regions. The modulation frequency is $\omega_{\mathrm p}/2\pi=20~\mathrm{THz}$.
}
\label{fig:exp_schem}
\end{figure*}

In this Letter, we show that Floquet engineering of Landau polaritons~\cite{PhysRevB.81.235303,Scalari2012,Zhang2016Collective,Li2018Observation}, formed by coupling the cyclotron resonance (CR) of a two-dimensional electron gas in a perpendicular magnetic field to a terahertz (THz) cavity mode (Fig.~\ref{fig:exp_schem}a), can circumvent the no-go theorem forbidding an SRPT in equilibrium. We consider a time-periodic magnetic modulation that drives both the cyclotron frequency and the light--matter coupling strength, while leaving the $\boldsymbol{A}^2$ term unchanged. In the high-frequency regime, the system exhibits an effective DC renormalization of the coupling strength analogous to optical rectification, enabling $\Omega_{\mathrm{eff}}>\sqrt{\omega_0 D}$ and thus access to a quantum critical point set by the modulation amplitude and static coupling strength. Beyond this point, the system enters a superradiant phase with macroscopic cavity-photon occupation and coherent amplitude, accompanied by in-plane electronic polarization arising from band nonparabolicity in the semiconductor host (Fig.~\ref{fig:exp_schem}a,b). In contrast to driven-dissipative scenarios, this quasiequilibrium approach operates far from resonance, such that the laser drive does not inject net energy into the system, thereby providing a new route to SRPTs. Finally, we show that this regime is accessible with realistic midinfrared picosecond modulation and predict photon bursts as a direct experimental signature.

We consider a two-dimensional spinless electron gas of area $S$ and density $\rho$, subjected to a perpendicular magnetic field $B_0$ and coupled to a single cavity mode of frequency $\omega_{\mathrm{cav}}$, effective thickness $L$, and $y$-polarization (Fig.~\ref{fig:exp_schem}a). In the Landau gauge, states are labeled by the Landau level (LL) index $n$ and momentum $k$ along $y$, with guiding-center position $k l_0^2$, where $l_0 = \sqrt{\hbar/eB}$ is the magnetic length. The degeneracy of each LL is $N=S/2\pi l_0^2$. Adjacent LLs are separated by the cyclotron frequency $\omega_0 = eB_0/m$, where $-e$ and $m$ denote the electron charge and effective mass, respectively; we assume resonance with the cavity mode, $\omega_0 = \omega_{\mathrm{cav}}$. The system is described by the Hamiltonian~\cite{PhysRevB.81.235303,Note1}
\begin{align}
\label{Hamiltonian_fermions}
\hat{H} & = \sum_{n,k}n \hbar \omega_{0} \hat{c}^\dagger_{n,k}\hat{c}_{n,k}+ \hbar\omega_{0}\hat{a}^\dagger \hat{a} +  \hbar D\left(\hat{a}+\hat{a}^\dagger\right)^2 \nonumber \\
& +\sum_{n,k} \hbar g_{0} \sqrt{\frac{n+1}{N}}\left(\hat{a}+\hat{a}^\dagger\right)\left(\hat{c}^\dagger_{n+1,k}\hat{c}_{n,k}+ \mathrm{h.c.}\right),
\end{align}
where $\hat{a}$ and $\hat{c}_{n,k}$ annihilate a cavity photon and an electron, respectively, and $g_0=\sqrt{\alpha_{\mathrm{fs}}\omega_0 c/L}$, with $\alpha_{\mathrm{fs}}$ the fine-structure constant and $c$ the speed of light in the intracavity medium. The diamagnetic coupling strength is $D=g_0^2\nu/\omega_0$, with $\nu=2\pi l_0^2\rho$ the LL filling factor; notably, $D$ is independent of the applied magnetic field.

Owing to Pauli blocking, the dynamics can be restricted to the two LLs adjacent to the Fermi energy, namely the highest occupied level $n=\nu-1$ and the lowest unoccupied level $n=\nu$. Introducing the collective CR excitation operator $\hat{b} = \frac{1}{\sqrt{N}} \sum_{k} 
\hat{c}^{\dagger}_{\nu-1,k}\hat{c}^{\vphantom{\dagger}}_{\nu,k}$ and $\hat{b}^{\dagger}$, which involve all guiding-center states, the Hamiltonian can be rewritten within this reduced Hilbert space, as shown in the Supplemental Material~\cite{Note1}. In the dilute-excitation regime, $\hat{b}$ and $\hat{b}^\dagger$ approximately obey bosonic commutation relations, and the Hamiltonian reduces to the Hopfield model~\cite{Hopfield1958} with collective coupling strength $\Omega_0=g_0\sqrt{\nu}$, yielding two polariton eigenmodes with frequencies $\omega_{\mathrm{LP}}$ and $\omega_{\mathrm{UP}}$. Crucially, the TRK sum rule imposes the relation $D=\Omega_0^2/\omega_0$~\cite{NatafCiuti2010}, which ties the diamagnetic and collective light--matter coupling strengths. As a consequence, the model does not exhibit a SRPT: The lower-polariton eigenfrequency
\begin{equation}
\label{lower_polariton}
\omega_{\mathrm{LP}}=\omega_0\sqrt{\left(1+\frac{2D}{\omega_0}\right)-\frac{2}{\omega_0}\sqrt{D^2+\Omega^2_{0}}},
\end{equation}
remains finite for any finite value of $\Omega_0$.

To overcome this constraint, our Landau polariton system is now subjected to a periodic magnetic field applied perpendicular to the plane~\cite{HagenmullerDavid2016AdCe}, such that the total magnetic field reads $B(t) = B_0 + \Delta B \cos(\omega_\mathrm{p} t)$, where $\Delta B$ and $\omega_\mathrm{p}$ denote the amplitude and angular frequency of the modulation, respectively. This periodic drive modulates the cyclotron frequency according to $\omega_{\mathrm{cyc}} (t) = \omega_0 \left[ 1 + \varepsilon \cos(\omega_\mathrm{p} t) \right]$, with $\varepsilon = \Delta B / B_0$ the relative modulation strength. Since the light--matter coupling scales as the square root of the cyclotron frequency, it becomes $g(t) = g_0 \sqrt{\left| 1 + \varepsilon \cos(\omega_\mathrm{p} t) \right|}$. The resulting dynamics is governed by the $T$-periodic Floquet Hamiltonian, $\hat H(t+T)=\hat H(t)$ with $T=2\pi/\omega_{\mathrm p}$,
\begin{align}
\label{eq:hopfield_model}
\hat{H}(t) &= \hbar\omega_{\mathrm{cyc}}(t) \hat{b}^\dagger \hat{b}
+ \hbar\omega_{0} \hat{a}^\dagger \hat{a}
+ \hbar D \left(\hat{a}^\dagger + \hat{a}\right)^2 \nonumber \\
& + \hbar \Omega(t) \left(\hat{a}^\dagger + \hat{a}\right)
\left(\hat{b}^\dagger + \hat{b}\right),
\end{align}
where $\Omega(t)=g(t)\sqrt{\nu}$. Equation~\eqref{eq:hopfield_model} retains the form of the Hopfield Hamiltonian, with the periodic drive modulating the cyclotron frequency and collective coupling strength but leaving the diamagnetic term unchanged. Note that such periodic modulation of light--matter interactions has also been recently explored in the ultrastrong-coupling regime of the quantum Rabi model~\cite{PhysRevLett.134.063602}. The onset of the Floquet-driven SRPT is directly evidenced by the dynamics generated by $\hat{H}(t)$, obtained by numerically solving the equations of motion for the CR and photon operators~\cite{Note1}. The dynamics over one period $T$ is characterized by the eigenvalues $e^{\mu_j T}$ of the one-period evolution matrix, where $\mu_j$ are the Floquet exponents. A dynamical instability occurs when the largest real part, $\mathrm{max}_j \mathrm{Re}\{\mu_j\}$, becomes positive, corresponding to exponential growth of the fields. In polaritonic systems, parametric instabilities can arise when the drive frequency lies within the polariton spectrum and its harmonics satisfy the resonance conditions for polariton-pair creation~\cite{DeLiberato2007}. Figure~\ref{fig:exp_schem}(c) shows $\max_j \mathrm{Re}\{\mu_j\}$ as a function of the equilibrium coupling strength $\Omega_0/\omega_0$ and relative modulation amplitude $\varepsilon$ in the high-frequency regime, where $\omega_{\mathrm p}$ exceeds all other energy scales. A region with $\max_j \mathrm{Re}\{\mu_j\}>0$ emerges above a threshold, signaling a Floquet instability.

To characterize this threshold and establish its connection to a SRPT, we employ a Floquet--Magnus expansion $\hat{H}^{\mathrm F}=\sum_{m=0}^{\infty}\hat{H}^{\mathrm F}_m$~\cite{Eckardt_2015}, controlled by the parameter $\varepsilon\omega_0/\omega_{\mathrm p}$. To leading order, $\hat{H}^{\mathrm{F}}_{0} = \frac{1}{T}\int_0^T \! dt \, \hat{H}(t)$, which governs the coarse-grained dynamics, while the higher-order terms $\hat{H}^{\mathrm F}_m=\mathcal{O}(1/\omega_{\mathrm p}^m)$ ($m\ge1$) describe virtual absorption and emission processes of drive quanta within one period. These corrections vanish as $\omega_{\mathrm p}\to\infty$, such that $\hat{H}^{\mathrm F}\simeq\hat{H}^{\mathrm F}_0$. Because $\omega_{\mathrm{cyc}}(t)$ depends linearly on the magnetic field, its time average is simply $\omega_0$. By contrast, the nonlinear dependence of the light--matter coupling on the magnetic field leads to the enhanced effective coupling
\begin{equation}
\label{eff_coupling}
\Omega_{\mathrm{eff}}
=\frac{\Omega_0}{T}\int_0^T dt\,\sqrt{\bigl|1+\varepsilon\cos(\omega_{\mathrm p}t)\bigr|}
\simeq \Omega_0\eta\sqrt{\varepsilon},
\end{equation}
where $\eta=\Gamma(3/4)/[\sqrt{\pi}\Gamma(5/4)]\approx0.762$ is a numerical prefactor of order unity. The leading-order Floquet Hamiltonian therefore takes the Hopfield form
\begin{align}
\label{eq:floquet_hamiltonian}
\hat{H}^{\mathrm F}_0
&=\hbar\omega_0\hat{b}^\dagger\hat{b}
+\hbar\omega_0\hat{a}^\dagger\hat{a}
+\hbar D(\hat{a}+\hat{a}^\dagger)^2 \nonumber\\
&\quad+\hbar\Omega_{\mathrm{eff}}
(\hat{a}^\dagger+\hat{a})
(\hat{b}^\dagger+\hat{b}),
\end{align}
but with a collective coupling $\Omega_{\mathrm{eff}}$ that explicitly depends on $\varepsilon$ and is no longer tied to $D$ through the TRK relation. Indeed, $\hat{H}^{\mathrm F}_0$ results from a time-averaging procedure and does not describe a static equilibrium system, unlike the underlying Hamiltonian in Eq.~\eqref{eq:hopfield_model} at every instant of time. Replacing $\Omega_0$ by $\Omega_{\mathrm{eff}}$ in Eq.~\eqref{lower_polariton}, we find that the lower-polariton eigenfrequency $\omega_{\mathrm{LP}}$ softens to zero when $\Omega_{\mathrm{eff}}$ reaches the critical value $\Omega_{\mathrm{c}}=\sqrt{\omega_0(\omega_0+4D)/4}$. This occurs when the modulation amplitude exceeds the threshold value $\varepsilon_{\mathrm c}$ defined by $\eta^2 \varepsilon_\mathrm{c} = 1+\omega_0^2/(4\Omega_0^2)$, shown as a white dashed line in Fig.~\ref{fig:exp_schem}(c). The threshold accurately delineates the onset of the Floquet-instability region and marks the transition to a superradiant phase with macroscopic cavity-photon occupation and electronic polarization.

Although the normal phase of our model is formally analogous to a Dicke model supplemented by an $\boldsymbol{A}^{2}$-term with $\Omega_{\mathrm{eff}} < \sqrt{\omega_{0}D}$, namely, a collection of two-level emitters coupled to a cavity mode, the superradiant phase exhibits crucial qualitative differences. In particular, the restriction to the LL subspace $n=\nu-1,\nu$ (two-level approximation) breaks down. The macroscopic cavity occupation is accompanied by a macroscopic population of LLs above the Fermi energy, comparable to that below it~\cite{Note1}, such that all LLs must, in principle, be included. Consequently, the bosonic description in terms of Eq.~\eqref{eq:floquet_hamiltonian} is no longer valid, as the CR operators $\hat{b}$ and $\hat{b}^\dagger$ cease to obey bosonic commutation relations in this regime. Thus, to describe the superradiant phase, we first re-express the collective operators $\hat{b}$ and $\hat{b}^\dagger$ in terms of fermionic operators. This yields an Hamiltonian analogous to Eq.~\eqref{Hamiltonian_fermions}, with a Floquet-renormalized coupling $g_{0}=\Omega_{0}/\sqrt{\nu} \rightarrow \Omega_{\mathrm{eff}}/\sqrt{\nu}$. 

We then adopt a standard mean-field approach, replacing photon operators by a real macroscopic field $\hat{a} \rightarrow \langle G (\alpha) \vert \hat{a} \vert G(\alpha)\rangle \equiv \alpha$ (and similarly for $\hat{a}^\dagger$), where $\vert G(\alpha)\rangle$ denotes the ground state. In the superradiant phase, the ground-state energy $E_\mathrm{G}(\alpha)$ exhibits two symmetric minima at $\pm \alpha_0 = \mathcal{O}(\sqrt{N})$. Photon condensation typically emerges in one of the minima through spontaneous parity-symmetry breaking (Fig.~\ref{fig:exp_schem}a).

Importantly, the mean-field Floquet Hamiltonian associated with an infinite ladder of equally spaced LLs does not support photon condensation~\cite{Note1}. As for any harmonic oscillator, the paramagnetic light--matter interaction shifts all LLs uniformly by $-4\hbar\alpha^2\Omega_{\mathrm{eff}}^2/N$, such that the ground-state energy remains strictly quadratic in the order parameter. Once its curvature changes sign at the critical point, the energy becomes unbounded from below, precluding a stable ordered phase without higher-order terms [i.e., $\mathcal{O}(\alpha^4)$]. This is a generic feature of symmetry-breaking phase transitions~\cite{LandauLifshitz1980}. Furthermore, introducing a rigid ultraviolet cutoff in the LL ladder yields cutoff-dependent order parameters and thus violates gauge invariance~\cite{Chirolli2012}. In realistic systems, band nonparabolicity naturally generates the higher-order terms required to stabilize the superradiant phase. We therefore incorporate the nonparabolicity of the GaAs conduction band as a physically motivated smooth cutoff. The resulting effective Floquet mean-field Hamiltonian reads $\overline{H^{\mathrm{F}}}=\hbar(\omega_{\mathrm{0}}+4D)\alpha^2 + \sum_k \overline{\mathcal{H}^{\mathrm{F}}_k}$, where
\begin{align}
\label{eq:non-parabolic-hamiltonian}
\overline{\mathcal{H}^{\mathrm{F}}_k} &= \sum_n \hbar \xi_n \hat{c}^\dagger_{n,k}\hat{c}_{n,k} \nonumber \\
&\quad + \sum_n 2\alpha \hbar \Omega_{\mathrm{eff}} \sqrt{\frac{n+1}{\nu N}} f_n 
\left(\hat{c}^\dagger_{n+1,k}\hat{c}_{n,k} + \mathrm{h.c.}\right),
\end{align}
where $\xi_n=\frac{\omega_0(n\omega_0-\gamma n^2)}{\omega_0-(2\nu-1)\gamma}$ and $f_n=\sqrt{\frac{\omega_0-(2n+1)\gamma}{\omega_0-(2\nu-1)\gamma}}$. The parameter $\gamma$ thus accounts for the reduction of LL spacing at large $n$ in GaAs~\cite{PFEFFERP1990CeiG,ZybertM.2017Llas}. 

\begin{figure*}[htbp]
    \centering
    \begin{minipage}{0.20\textwidth}
        \includegraphics[width=\linewidth]{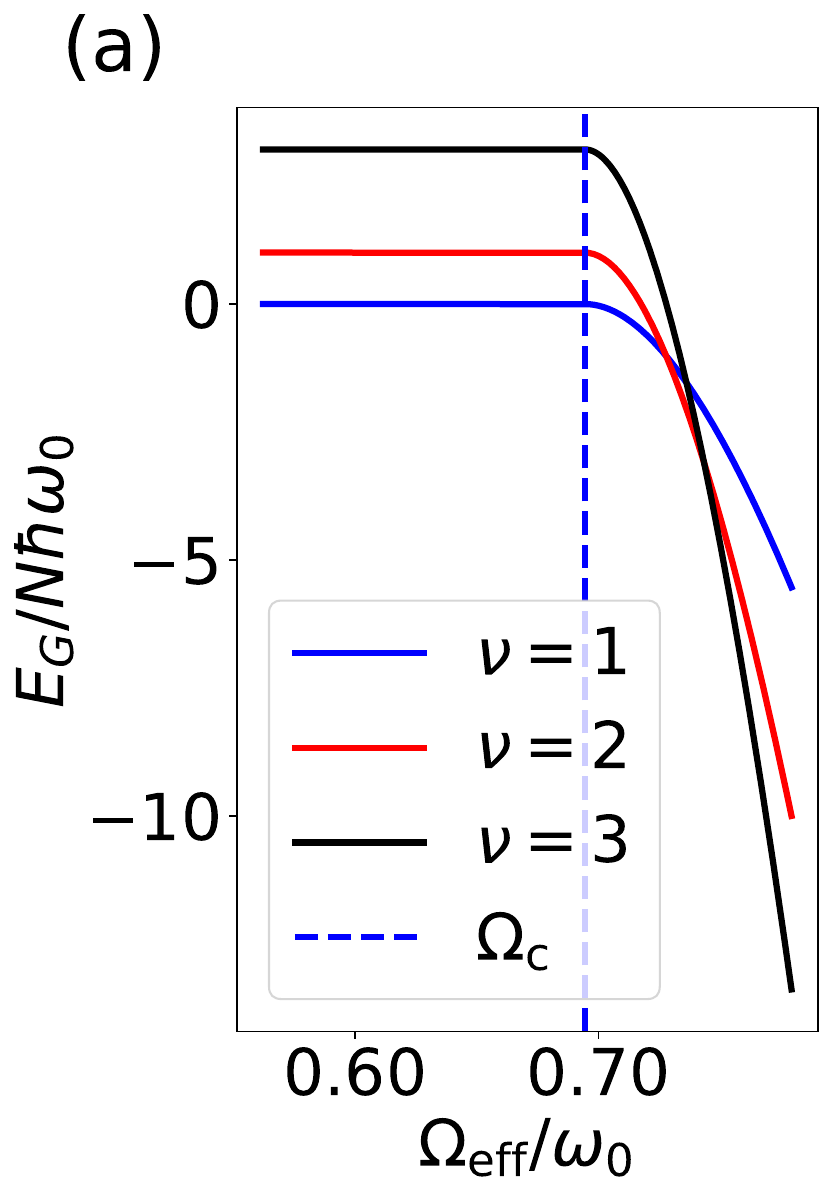}
    \end{minipage}
    \begin{minipage}{0.28\textwidth}
        \begin{overpic}[width=\linewidth]{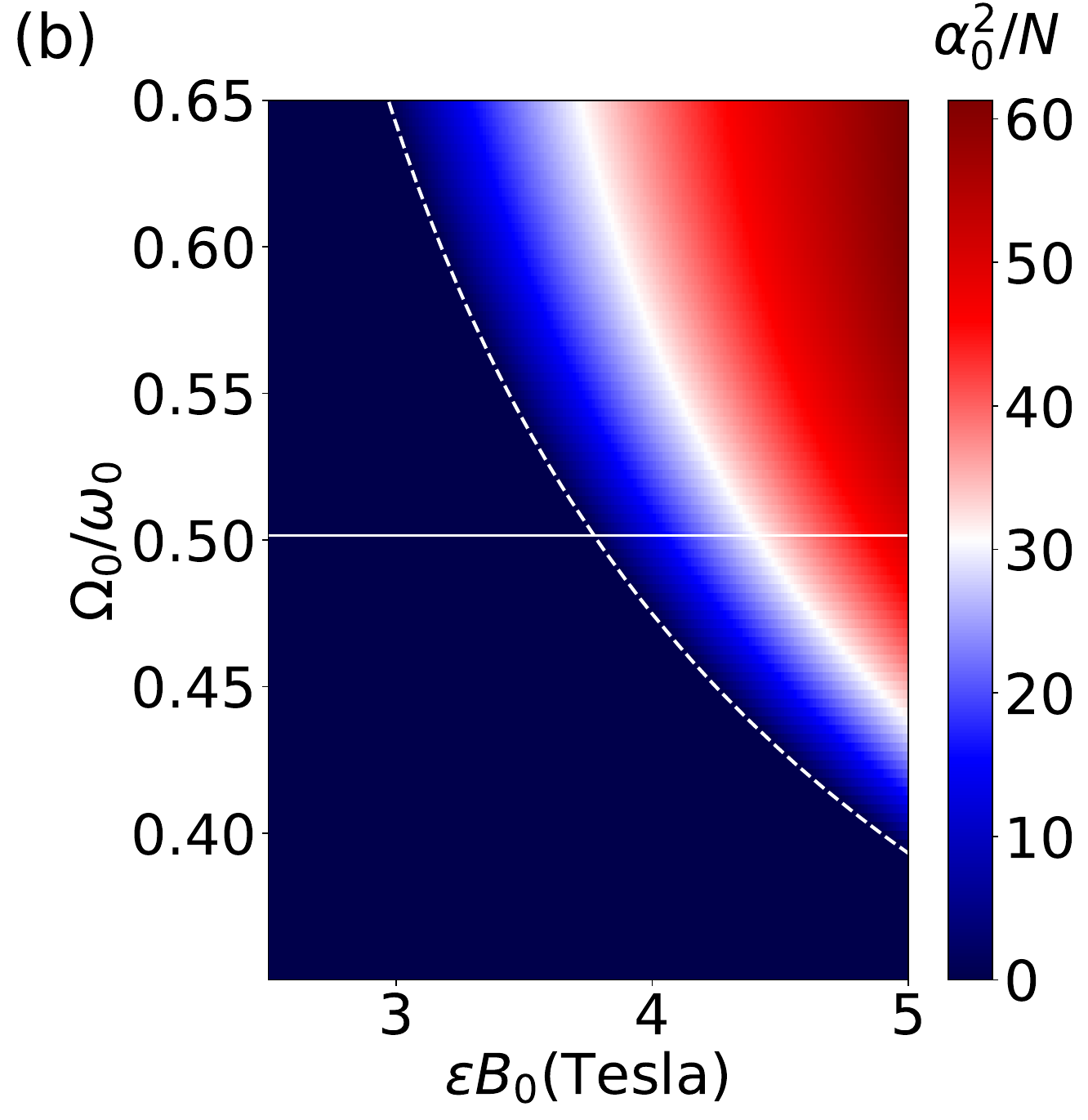}
            \put(25,14){\includegraphics[width=0.32\linewidth]{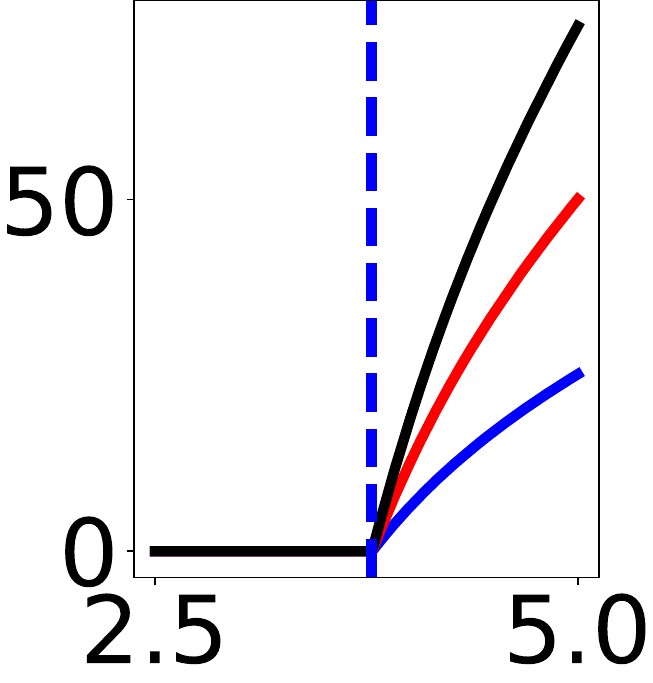}}
        \end{overpic}
    \end{minipage}
    \begin{minipage}{0.28\textwidth}
        \begin{overpic}[width=\linewidth]{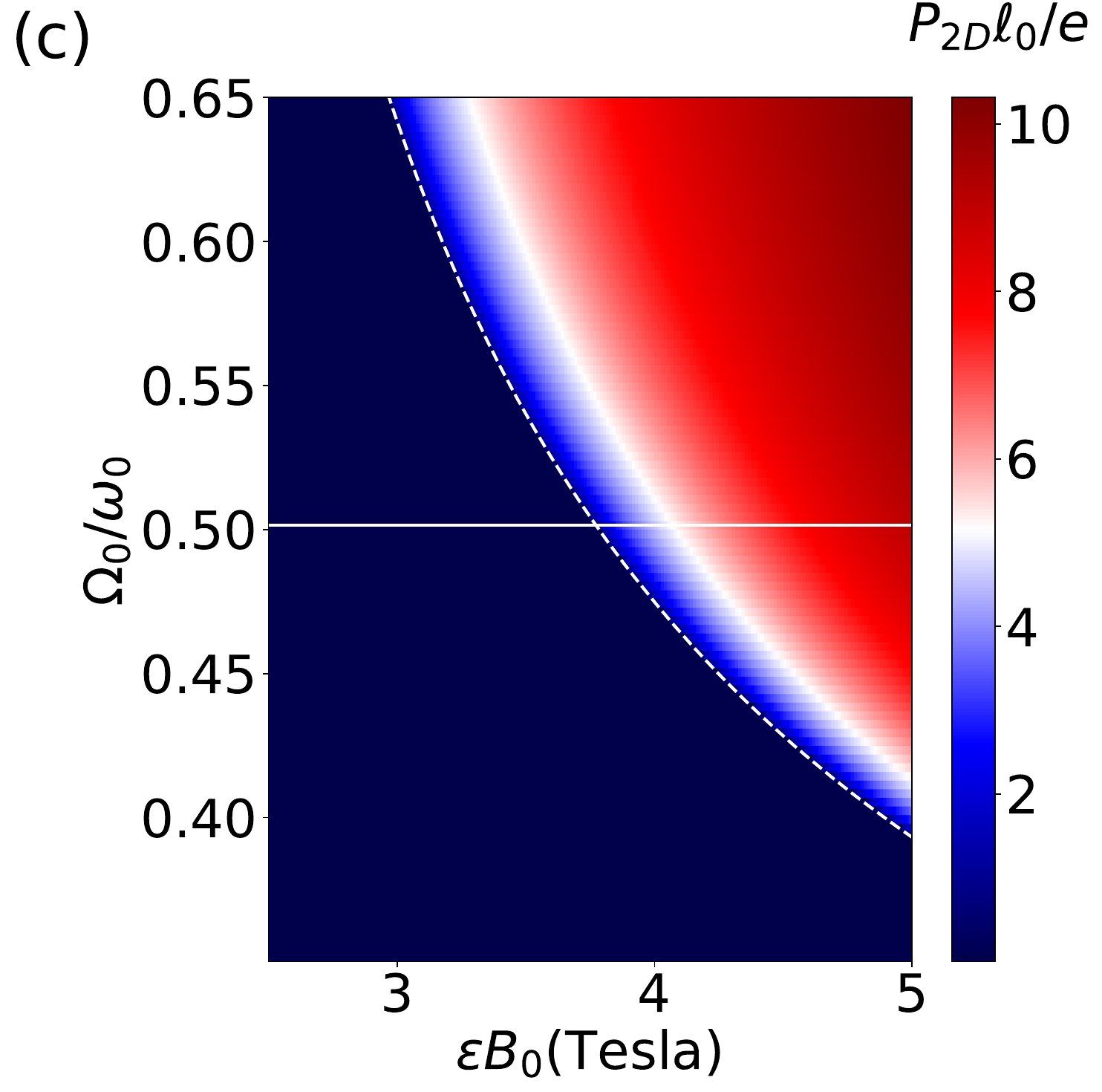}
            \put(25,13){\includegraphics[width=0.30\linewidth]{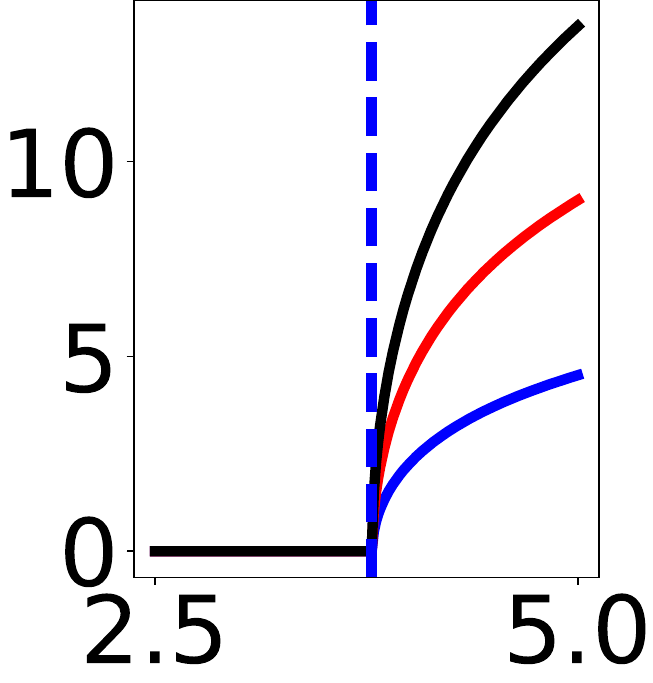}}
        \end{overpic}
    \end{minipage}
    \begin{minipage}{0.22\textwidth}
        \begin{overpic}[width=\linewidth]{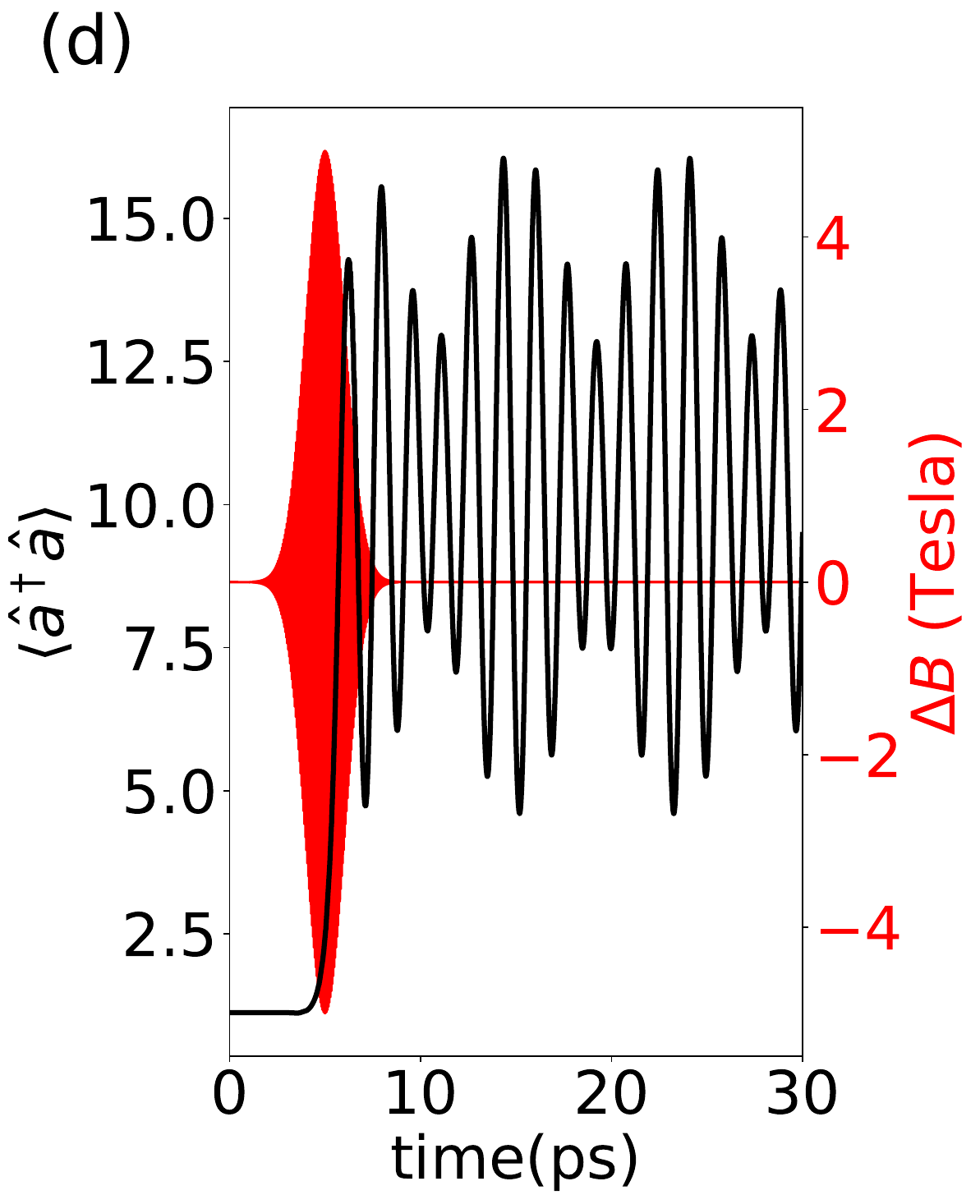}
            \put(34, 13){\includegraphics[width=0.5\linewidth]{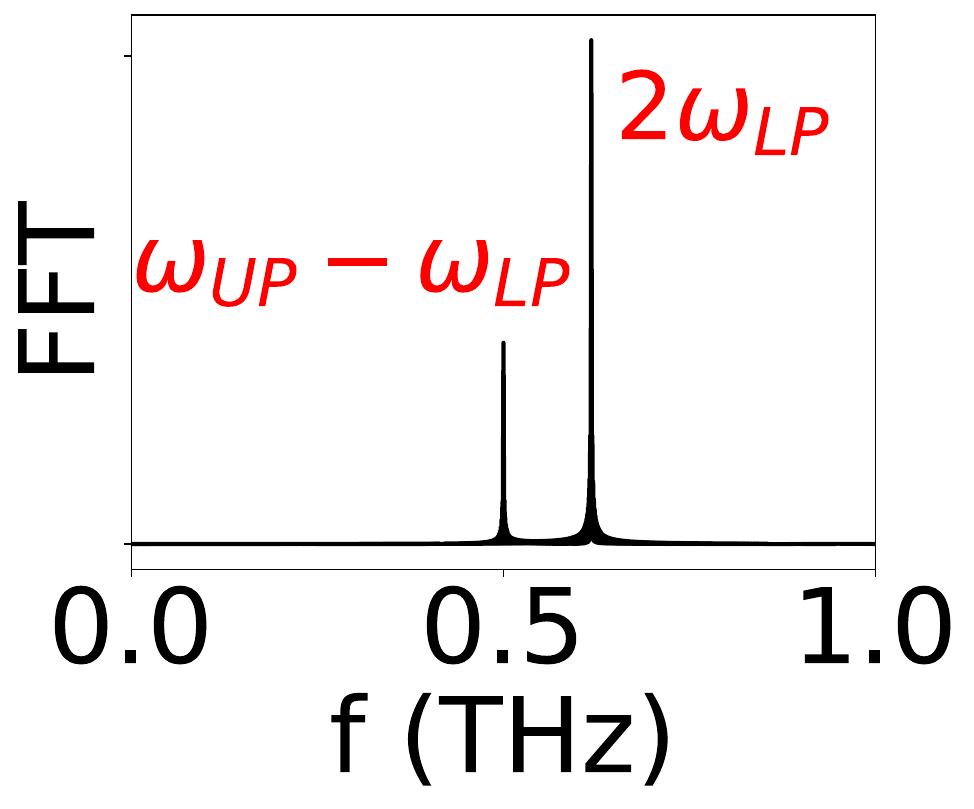}}
        \end{overpic}
    \end{minipage}
    \caption{\textbf{Critical behavior of the Floquet SRPT.} 
    (a)~Rescaled ground-state energy $E_\mathrm{G}(\alpha)/(N\hbar \omega_{0})$ versus $\Omega_{\mathrm{eff}}/\omega_{0}$ (evaluated at the corresponding field value $\alpha$) for different filling factors $\nu$. The blue dashed line marks the critical point $\Omega_{\mathrm{eff}}=\Omega_\mathrm{c}$. 
    (b)~Rescaled squared photon coherence $\alpha_0^2/N$ versus $\Omega_{0}/\omega_{0}$ and $\varepsilon B_0$ for $\nu=2$ (carrier density $\rho=5\times10^{10}$cm$^{-2}$). 
    (c)~Normalized sheet polarization $P_{\mathrm{2D}}\ell_0/e$ versus $\Omega_{0}/\omega_{0}$ and $\varepsilon B_0$ for $\nu=2$. The phase boundary $\eta^2 \varepsilon_\mathrm{c} = 1+\omega_0^2/(4\Omega_0^2)$ (see text) is shown as white dashed lines in (b) and (c); insets display horizontal cuts at $\Omega_0/\omega_0 = 0.5$. Blue, red, and black curves correspond to filling factors $\nu = 1, 2,$ and $3$, respectively, while dashed blue lines mark the critical amplitude $\varepsilon = \varepsilon_\mathrm{c}$. 
     (d)~Intracavity photon number $\langle \hat{a}^\dagger \hat{a} \rangle$ as a function of time for $\varepsilon B_0=5~\mathrm{T}$ and $\Omega_0/\omega_0=0.5$. Inset: Fourier spectrum of $\langle \hat{a}^\dagger \hat{a} \rangle(t)$ versus frequency $f=\omega/(2\pi)$. The low-frequency peak corresponds to the Rabi splitting, $\omega_{\mathrm{UP}}-\omega_{\mathrm{LP}}=2\Omega_0$, while the high-frequency peak appears at twice the lower-polariton frequency, $2\omega_{\mathrm{LP}}$. The latter originates from the softening of the lower-polariton branch, as discussed in the main text. Parameters: $\omega_{0}/(2\pi)=0.5$ THz, $\gamma=0.002\,\omega_0$, corresponding to an effective mass renormalization $m^*(n)=m^*(0)\left(1+n\hbar\omega_0/\epsilon_\mathrm{g}^*\right)$, with the effective energy gap of the two-band model $\epsilon_\mathrm{g}^* = 0.98\,\mathrm{eV}$~\cite{PFEFFERP1990CeiG,ZybertM.2017Llas}. All quantities are obtained from numerical diagonalization of the electronic Hamiltonian, as explained in the main text.}
    \label{fig:orderparam_2d}
\end{figure*}

The ground-state energy of $\overline{H^{\mathrm{F}}}$ factorizes as 
\begin{equation}
\label{ground_state_formula}
E_\mathrm{G}(\alpha)=\hbar(\omega_{\mathrm{0}}+4D)\alpha^2 + N \mathcal{E}_\mathrm{G}(\alpha),
\end{equation}
where $\mathcal{E}_\mathrm{G}(\alpha)=\langle G(\alpha)|\overline{\mathcal{H}^{\mathrm{F}}_k}|G(\alpha)\rangle$ denotes the electronic contribution to the ground-state energy, independent of $k$. It is instructive to perform a perturbative expansion of $\mathcal{E}_\mathrm{G}(\alpha)$ to second order in $\Omega_{\mathrm{eff}}$ (and $\alpha$). This yields $E_\mathrm{G}(\alpha) = \hbar\left[\omega_0 + 4D - (4\Omega_{\mathrm{eff}}^2/\omega_0)\right]\alpha^2 + \mathcal{O}(\alpha^4)$, up to a positive constant term~\cite{Note1}. The curvature at $\alpha=0$ becomes negative for $\Omega_{\mathrm{eff}}>\Omega_\mathrm{c} \equiv \sqrt{\omega_0(\omega_0+4D)/4}$, thereby identifying the quantum critical point, in agreement with the lower-polariton softening condition derived above. We then diagonalize the electronic Hamiltonian $\overline{\mathcal{H}^{\mathrm{F}}_k}$ numerically in the LL basis and compute its ground-state energy $\mathcal{E}_\mathrm{G}(\alpha)$, which is subsequently used in Eq.~\eqref{ground_state_formula} to obtain the full energy $E_\mathrm{G}(\alpha)$ including the photonic contribution. The resulting $E_G(\alpha)$ is shown in Fig.~2a as a function of the effective coupling ratio $\Omega_{\mathrm{eff}}/\omega_{0}$. It evolves from a constant value in the normal phase and decreases in the superradiant phase, in agreement with the perturbative analysis above. The photonic order parameter $\alpha_0$ is obtained by minimizing $E_\mathrm{G}(\alpha)$ with respect to $\alpha$, i.e., $\alpha_0 = \arg\min_{\alpha} E_\mathrm{G}(\alpha)$. 

A natural order parameter for the matter sector is the static in-plane polarization, which is associated with an instability towards dipolar transitions between LLs, akin to a ferroelectric response. By recasting the ground-state expectation value of the light--matter coupling term in $\overline{H^{\mathrm{F}}}$ as $-d \cdot E$, with $E=\alpha\sqrt{\hbar\omega_0/(\epsilon_0 V)}$ the macroscopic electric field ($V=SL$ and $\epsilon_0$ the vacuum permittivity), we identify the static collective in-plane dipole moment as
\begin{align*}
d = - d_0 \sum_{n} \sqrt{n+1} f_{n} \Re{\langle G(\alpha)| \hat{c}^\dagger_{n+1,k}\hat{c}_{n,k}|G(\alpha)\rangle},
\end{align*}
where $d_0=eS \eta \sqrt{2\varepsilon}/(\pi l_{0})$. The corresponding sheet polarization is $P_{\mathrm{2D}} = d/S$, evaluated from the ground state $\vert G(\alpha)\rangle$ obtained by numerical diagonalization of $\overline{\mathcal{H}^{\mathrm{F}}_k}$. The rescaled order parameters $\alpha_0^2/N$ and $P_{\mathrm{2D}}\ell_0/e$ are plotted as functions of $\Omega_0/\omega_0$ and $\varepsilon B_0$ in Figs.~\ref{fig:orderparam_2d}b,c. Both quantities continuously increase beyond the critical point defined by $\Omega_{\mathrm{eff}}=\Omega_\mathrm{c}$, displaying the characteristic behavior of a Dicke-like second-order phase transition. This is confirmed by a perturbative expansion up to fourth order in $\alpha$, yielding $\alpha_0 \sim\sqrt{\Omega_{\mathrm{eff}}^2-\Omega_\mathrm{c}^2}$ in the vicinity of the critical point. 

To establish a direct correspondence with the Dicke model, we also introduce the excitation number
\begin{equation}
\label{eq:beta_0}
\beta_0^2 = \sum_{n,k} n \left(\langle G(\alpha)|\hat{c}^\dagger_{n,k}\hat{c}_{n,k}|G(\alpha)\rangle - \langle G^{(0)}|\hat{c}^\dagger_{n,k}\hat{c}_{n,k}|G^{(0)}\rangle \right),
\end{equation}
where $|G^{(0)}\rangle$ denotes the ground state in the absence of light--matter coupling, with the LLs filled up to $n=\nu-1$. This quantity measures the excess number of cyclotron quanta stored in the LLs due to the light--matter interaction. It is analogous to the deviation of the collective spin polarization $\langle J_z\rangle$ from its fully polarized value $-N/2$ in the Dicke model~\cite{Note1,EmaryBrandes2003}. Since the electronic mean-field Hamiltonian decomposes into independent contributions for each $k$, one finds $\beta_0=\mathcal{O}(\sqrt{N})$. Moreover, close to the critical point, $\beta_0 \sim\sqrt{\Omega_{\mathrm{eff}}^2-\Omega_\mathrm{c}^2}$
in agreement with the critical scaling of the excitation number in the Dicke model~\cite{Note1}.

Reaching the superradiant phase requires magnetic-field modulations of several tesla, raising questions of experimental feasibility. Recent microstructured devices, such as spiral antennas, can generate picosecond AC magnetic fields normal to the sample with amplitudes 10--100 times larger than the incident pump~\cite{ZhangZhenya2025SsiS}. With a mid-infrared drive, this enables transient modulations of order $\varepsilon B_0 \sim 1$--10~T. To assess whether such pulses induce observable signatures of photon condensation and in-plane ferroelectricity, we consider a Gaussian modulation $\Delta B(t) = \varepsilon B_0 \exp\left(-t^2/2t_\mathrm{p}^2 \right)\cos \left(\omega_\mathrm{p} t \right)$, with $t_\mathrm{p} = 1$~ps and $\omega_\mathrm{p}/2\pi=20$~THz, and compute the dynamics from the time-dependent Hamiltonian $\hat{H}(t)$. The resulting intracavity photon population $\langle \hat{a}^\dagger \hat{a} \rangle (t)$ is shown in Fig.~\ref{fig:orderparam_2d}d. Following the pulse, the photon population exhibits beatings combining Rabi oscillations at $2\Omega_0=\omega_{\mathrm{UP}}-\omega_{\mathrm{LP}}$ with a high-frequency component at $2\omega_{\mathrm{LP}}$. The latter directly reflects the Floquet-induced superradiant parametric instability through the softening of the lower-polariton mode. Averaging over several Rabi periods, we find that the intracavity photon population increases by several orders of magnitude for $\Omega_0>\Omega_\mathrm{c}$ and $\varepsilon>\varepsilon_\mathrm{c}$~\cite{Note1}, although this estimate becomes unreliable beyond the critical point, where the normal-phase description breaks down. 

Physically, the pulse quenches the system from the dressed vacuum corresponding to the bare coupling $\Omega_0$ into a Floquet quasiequilibrium state characterized by the enhanced coupling $\Omega_{\mathrm{eff}}$. In the presence of cavity losses, this quench is expected to generate a coherent burst of photons emitted from the cavity, providing a direct dynamical signature of Floquet-induced photon condensation before the system relaxes back to equilibrium~\cite{Ciuti2005,HagenmullerDavid2016AdCe}. The number of emitted photons can be estimated from the change in ground-state energy, $N_{\mathrm{burst}} \sim \vert E_\mathrm{G}(\Omega^{\mathrm{max}}_{\mathrm{eff}})-E_\mathrm{G}(\Omega_{0})\vert/\hbar\omega_0$, where $\Omega^{\mathrm{max}}_{\mathrm{eff}}=\Omega_0 \eta \sqrt{\Delta B/B_0}$ with $\Delta B$ taken at the peak of the Gaussian envelope. Since $E_\mathrm{G} = \mathcal{O}(N)$, $N_{\mathrm{burst}}$ also scales with the LL degeneracy.

In realistic systems, degeneracy scales with the effective cavity mode area, so the light--matter coupling strength stays independent of the surface. We consider $\varepsilon B_0 = 5\,\mathrm{T}$ and $\omega_0 / 2\pi = 0.5\,\mathrm{THz}$, corresponding to $B_0 \approx 1\,\mathrm{T}$ and a wavelength $\lambda = 600\,\upmu\mathrm{m}$. For metasurface resonators~\cite{Scalari2012,Bayer2017}, a typical mode extent $L \sim 0.01\lambda$ yields a coupling strength $\Omega_0 \approx 0.5\,\omega_0$. Combined with the cavity mode area $S \sim 20\,\upmu\mathrm{m}^{2}$, this results in $N_{\mathrm{burst}} \approx 6 \times 10^{4}$ and $P_{\mathrm{2D}} \approx 5.7 \times 10^{-11}\,\mathrm{C\,m^{-1}}$. Note that, for these experimental parameters, the leading-order Floquet Hamiltonian, Eq.~\eqref{eq:floquet_hamiltonian}, provides an excellent description of the dynamics, as the expansion parameter controlling the Floquet--Magnus series is small, i.e., $\varepsilon \omega_0/\omega_\mathrm{p} \simeq 0.1$.

In conclusion, we have shown that high-frequency Floquet engineering of a Landau polariton system enables a quasiequilibrium SRPT. Beyond the critical value of modulation amplitude, the system exhibits macroscopic cavity occupation and in-plane electronic polarization set by band nonparabolicity. This lossless, off-resonant scheme provides a route to SRPTs, bridging equilibrium settings, where no-go theorems or stringent constraints apply, and driven-dissipative scenarios that are inherently out of equilibrium and rely crucially on losses. Our results further open access to the parametric regime, where driving within the polariton spectrum can induce dynamical instabilities and driven-dissipative phase transitions in Landau polariton systems.

\footnotetext[1]{See Supplemental Material for details of the theoretical framework and numerical methods, including the derivation of the equilibrium Hamiltonian, the dynamics of the time-dependent Hamiltonian, the breakdown of the two-level approximation in the superradiant phase, the analogy with the matter order parameter of the Dicke model, the absence of an SRPT in an infinite LL ladder, and the unphysical nature of a rigid cutoff. The Supplemental Material includes Refs.~\cite{Ciuti2005,Chirolli2012}.}

\begin{acknowledgments}
\textit{Acknowledgments}--J.K.\ acknowledges support from the U.S.\ Army Research Office (through Award No.\ W911NF-25-2-0150, W911NF-21-1-0157), the Gordon and Betty Moore Foundation (through Grant No.\ 11520), and the Robert A.\ Welch Foundation (through Grant No.\ C-1509). 
\end{acknowledgments}

\bibliography{references}

@book{LandauLifshitz1980,
  author    = {L. D. Landau and E. M. Lifshitz},
  title     = {Statistical Physics, Part 1},
  series    = {Course of Theoretical Physics},
  volume    = {5},
  edition   = {3},
  publisher = {Butterworth-Heinemann},
  address   = {Oxford},
  year      = {1980},
  isbn      = {978-0-08-057046-4},
  url       = {https://shop.elsevier.com/books/statistical-physics/landau/978-0-08-057046-4}
}

@article{Baumann2010,
  author       = {Baumann, K. and Guerlin, C. and Brennecke, F. and Esslinger, T.},
  title        = {{D}icke Quantum Phase Transition with a Superfluid Gas in an Optical Cavity},
  journal      = {Nature},
  volume       = {464},
  number       = {7293},
  pages        = {1301--1306},
  year         = {2010},
  publisher    = {Nature Publishing Group},
  doi          = {10.1038/nature09009}
}

@article{NatafCiuti2010,
  author       = {Nataf, Pierre and Ciuti, Cristiano},
  title        = {No-go theorem for superradiant quantum phase transitions in cavity {QED} and counter-example in circuit {QED}},
  journal      = {Nat. Commun.},
  volume       = {1},
  pages        = {72},
  year         = {2010},
  doi          = {10.1038/ncomms1069}
}

@article{RousseauFelbacq2017,
  author       = {Rousseau, Emmanuel and Felbacq, Didier},
  title        = {The quantum-optics Hamiltonian in the Multipolar gauge},
  journal      = {Sci. Rep.},
  volume       = {7},
  pages        = {11115},
  year         = {2017},
  publisher    = {Nature Publishing Group},
  doi          = {10.1038/s41598-017-11076-5}
}

@article{KnightAharonovHsieh1978,
  author       = {Knight, J. M. and Aharonov, Y. and Hsieh, G. T. C.},
  title        = {Are super-radiant phase transitions possible?},
  journal      = {Phys. Rev. A},
  volume       = {17},
  number       = {4},
  pages        = {1454--1462},
  year         = {1978},
  publisher    = {American Physical Society},
  doi          = {10.1103/PhysRevA.17.1454}
}

@article{Andolina2020,
  author       = {Andolina, G. M. and Pellegrino, F. M. D. and Giovannetti, V. and MacDonald, A. H. and Polini, M.},
  title        = {Theory of photon condensation in a spatially varying electromagnetic field},
  journal      = {Phys. Rev. B},
  volume       = {102},
  number       = {12},
  pages        = {125137},
  year         = {2020},
  publisher    = {American Physical Society},
  doi          = {10.1103/PhysRevB.102.125137}
}

@article{Li2022,
  title = {Effective theory of lattice electrons strongly coupled to quantum electromagnetic fields},
  author = {Li, Jiajun and Schamri\ss{}, Lukas and Eckstein, Martin},
  journal = {Phys. Rev. B},
  volume = {105},
  issue = {16},
  pages = {165121},
  numpages = {16},
  year = {2022},
  month = {Apr},
  publisher = {American Physical Society},
  doi = {10.1103/PhysRevB.105.165121},
  url = {https://link.aps.org/doi/10.1103/PhysRevB.105.165121}
}

@article{Li2020,
  title = {Electromagnetic coupling in tight-binding models for strongly correlated light and matter},
  author = {Li, Jiajun and Golez, Denis and Mazza, Giacomo and Millis, Andrew J. and Georges, Antoine and Eckstein, Martin},
  journal = {Phys. Rev. B},
  volume = {101},
  issue = {20},
  pages = {205140},
  numpages = {16},
  year = {2020},
  month = {May},
  publisher = {American Physical Society},
  doi = {10.1103/PhysRevB.101.205140},
  url = {https://link.aps.org/doi/10.1103/PhysRevB.101.205140}
}

@article{Hopfield1958,
  title = {Theory of the Contribution of Excitons to the Complex Dielectric Constant of Crystals},
  author = {Hopfield, J. J.},
  journal = {Phys. Rev.},
  volume = {112},
  issue = {5},
  pages = {1555--1567},
  numpages = {0},
  year = {1958},
  month = {Dec},
  publisher = {American Physical Society},
  doi = {10.1103/PhysRev.112.1555},
  url = {https://link.aps.org/doi/10.1103/PhysRev.112.1555}
}

@Article{RomanRoche2022,
	title={{Effective theory for matter in non-perturbative cavity {QED}}},
	author={Juan Rom\'an-Roche and David Zueco},
	journal={SciPost Phys. Lect. Notes},
	pages={50},
	year={2022},
	publisher={SciPost},
	doi={10.21468/SciPostPhysLectNotes.50},
	url={https://scipost.org/10.21468/SciPostPhysLectNotes.50}
}

@article{Garziano2020,
  title = {Gauge invariance of the {D}icke and Hopfield models},
  author = {Garziano, Luigi and Settineri, Alessio and Di Stefano, Omar and Savasta, Salvatore and Nori, Franco},
  journal = {Phys. Rev. A},
  volume = {102},
  issue = {2},
  pages = {023718},
  numpages = {11},
  year = {2020},
  month = {Aug},
  publisher = {American Physical Society},
  doi = {10.1103/PhysRevA.102.023718},
  url = {https://link.aps.org/doi/10.1103/PhysRevA.102.023718}
}

@article{DiStefano2019,
  author       = {Di Stefano, Omar and Settineri, Alessio and Macr\`i, Vincenzo and Garziano, Luca and Stassi, Rosario and Savasta, Salvatore and Nori, Franco},
  title        = {Resolution of gauge ambiguities in ultrastrong-coupling cavity {QED}},
  journal      = {Nat. Phys.},
  volume       = {15},
  number       = {8},
  pages        = {803--808},
  year         = {2019},
  publisher    = {Nature Publishing Group},
  doi          = {10.1038/s41567-019-0534-4}
}

@article{DeBernardis2018,
  author       = {De Bernardis, Daniele and Pilar, Philipp and Jaako, Tuomas and De Liberato, Simone and Rabl, Peter},
  title        = {Breakdown of gauge invariance in ultrastrong-coupling cavity {QED}},
  journal      = {Phys. Rev. A},
  volume       = {98},
  number       = {5},
  pages        = {053819},
  year         = {2018},
  publisher    = {American Physical Society},
  doi          = {10.1103/PhysRevA.98.053819}
}

@article{Viehmann2011,
  title = {Superradiant Phase Transitions and the Standard Description of Circuit {QED}},
  author = {Viehmann, Oliver and von Delft, Jan and Marquardt, Florian},
  journal = {Phys. Rev. Lett.},
  volume = {107},
  issue = {11},
  pages = {113602},
  numpages = {5},
  year = {2011},
  month = {Sep},
  publisher = {American Physical Society},
  doi = {10.1103/PhysRevLett.107.113602},
  url = {https://link.aps.org/doi/10.1103/PhysRevLett.107.113602}
}

@article{Baksic2013,
  title = {Superradiant phase transitions with three-level systems},
  author = {Baksic, Alexandre and Nataf, Pierre and Ciuti, Cristiano},
  journal = {Phys. Rev. A},
  volume = {87},
  issue = {2},
  pages = {023813},
  numpages = {5},
  year = {2013},
  month = {Feb},
  publisher = {American Physical Society},
  doi = {10.1103/PhysRevA.87.023813},
  url = {https://link.aps.org/doi/10.1103/PhysRevA.87.023813}
}

@article{Hayn2011,
  title = {Phase transitions and dark-state physics in two-color superradiance},
  author = {Hayn, Mathias and Emary, Clive and Brandes, Tobias},
  journal = {Phys. Rev. A},
  volume = {84},
  issue = {5},
  pages = {053856},
  numpages = {8},
  year = {2011},
  month = {Nov},
  publisher = {American Physical Society},
  doi = {10.1103/PhysRevA.84.053856},
  url = {https://link.aps.org/doi/10.1103/PhysRevA.84.053856}
}

@article{Savasta2021,
url = {https://doi.org/10.1515/nanoph-2020-0433},
title = {{Thomas}--{Reiche}--{Kuhn} ({TRK}) sum rule for interacting photons},
author = {Salvatore Savasta and Omar Di Stefano and Franco Nori},
pages = {465--476},
volume = {10},
number = {1},
journal = {Nanophotonics},
doi = {10.1515/nanoph-2020-0433},
year = {2021},
lastchecked = {2026-01-30}
}

@article{Tufarelli2015,
  title = {Signatures of the ${A}^{2}$ term in ultrastrongly coupled oscillators},
  author = {Tufarelli, Tommaso and McEnery, K. R. and Maier, S. A. and Kim, M. S.},
  journal = {Phys. Rev. A},
  volume = {91},
  issue = {6},
  pages = {063840},
  numpages = {11},
  year = {2015},
  month = {Jun},
  publisher = {American Physical Society},
  doi = {10.1103/PhysRevA.91.063840},
  url = {https://link.aps.org/doi/10.1103/PhysRevA.91.063840}
}

@article{Vukics2015,
  title = {Fundamental limitation of ultrastrong coupling between light and atoms},
  author = {Vukics, Andr\'as and Grie\ss{}er, Tobias and Domokos, Peter},
  journal = {Phys. Rev. A},
  volume = {92},
  issue = {4},
  pages = {043835},
  numpages = {6},
  year = {2015},
  month = {Oct},
  publisher = {American Physical Society},
  doi = {10.1103/PhysRevA.92.043835},
  url = {https://link.aps.org/doi/10.1103/PhysRevA.92.043835}
}

@article{Bamba2016,
  title = {Superradiant Phase Transition in a Superconducting Circuit in Thermal Equilibrium},
  author = {Bamba, Motoaki and Inomata, Kunihiro and Nakamura, Yasunobu},
  journal = {Phys. Rev. Lett.},
  volume = {117},
  issue = {17},
  pages = {173601},
  numpages = {5},
  year = {2016},
  month = {Oct},
  publisher = {American Physical Society},
  doi = {10.1103/PhysRevLett.117.173601},
  url = {https://link.aps.org/doi/10.1103/PhysRevLett.117.173601}
}

@article{EmaryPRL2003,
  title = {Quantum Chaos Triggered by Precursors of a Quantum Phase Transition: The {{Dicke}} Model},
  author = {Emary, Clive and Brandes, Tobias},
  journal = {Phys. Rev. Lett.},
  volume = {90},
  issue = {4},
  pages = {044101},
  numpages = {4},
  year = {2003},
  month = {Jan},
  publisher = {American Physical Society},
  doi = {10.1103/PhysRevLett.90.044101},
  url = {https://link.aps.org/doi/10.1103/PhysRevLett.90.044101}
}

@article{Andolina2022NonPerturbative,
  author       = {Gian Marcello Andolina and Francesco M. D. Pellegrino and Alberto Mercurio and Oreste Di Stefano and Marco Polini and Salvatore Savasta},
  title        = {A non-perturbative no-go theorem for photon condensation in approximate models},
  journal      = {Eur. Phys. J. Plus},
  volume       = {137},
  pages        = {1348},
  year         = {2022},
  doi          = {10.1140/epjp/s13360-022-03571-0},
  url          = {https://doi.org/10.1140/epjp/s13360-022-03571-0}
}

@article{Kirton2019,
author = {Kirton, Peter and Roses, Mor M. and Keeling, Jonathan and Dalla Torre, Emanuele G.},
title = {Introduction to the {{Dicke}} Model: From Equilibrium to Nonequilibrium, and Vice Versa},
journal = {Adv. Quantum Technol.},
volume = {2},
number = {1-2},
pages = {1970013},
doi = {10.1002/qute.201970013},
url = {https://advanced.onlinelibrary.wiley.com/doi/abs/10.1002/qute.201970013},
year = {2019}
}

@article{NatafBasko2019,
  author       = {Nataf, Pierre and Champel, Thierry and Blatter, Gianni and Basko, Denis M.},
  title        = {Rashba Cavity {QED}: A Route Towards the Superradiant Quantum Phase Transition},
  journal      = {Phys. Rev. Lett.},
  volume       = {123},
  number       = {20},
  pages        = {207402},
  year         = {2019},
  publisher    = {American Physical Society},
  doi          = {10.1103/PhysRevLett.123.207402}
}

@article{Manzanares2022,
  author       = {Manzanares, Guillaume and Champel, Thierry and Basko, Denis M. and Nataf, Pierre},
  title        = {Superradiant quantum phase transition for Landau polaritons with Rashba and Zeeman couplings},
  journal      = {Phys. Rev. B},
  volume       = {105},
  number       = {24},
  pages        = {245304},
  year         = {2022},
  publisher    = {American Physical Society},
  doi          = {10.1103/PhysRevB.105.245304}
}

@article{DeBernardis2018_2,
  author       = {De Bernardis, Daniele and Jaako, Tuomas and Rabl, Peter},
  title        = {Cavity quantum electrodynamics in the nonperturbative regime},
  journal      = {Phys. Rev. A},
  volume       = {97},
  number       = {4},
  pages        = {043820},
  year         = {2018},
  publisher    = {American Physical Society},
  doi          = {10.1103/PhysRevA.97.043820}
}

@article{Guerci2020,
  author       = {Guerci, Daniele and Simon, Pascal and Mora, Christophe},
  title        = {Superradiant Phase Transition in Electronic Systems and Emergent Topological Phases},
  journal      = {Phys. Rev. Lett.},
  volume       = {125},
  number       = {25},
  pages        = {257604},
  year         = {2020},
  publisher    = {American Physical Society},
  doi          = {10.1103/PhysRevLett.125.257604}
}

@article{Dmytruk2021,
  author       = {Dmytruk, Olesia and Schir\'o, Marco},
  title        = {Gauge fixing for strongly correlated electrons coupled to quantum light},
  journal      = {Phys. Rev. B},
  volume       = {103},
  number       = {7},
  pages        = {075131},
  year         = {2021},
  publisher    = {American Physical Society},
  doi          = {10.1103/PhysRevB.103.075131}
}

@article{Abedinpour2011,
  title = {Drude weight, plasmon dispersion, and ac conductivity in doped graphene sheets},
  author = {Abedinpour, Saeed H. and Vignale, G. and Principi, A. and Polini, Marco and Tse, Wang-Kong and MacDonald, A. H.},
  journal = {Phys. Rev. B},
  volume = {84},
  issue = {4},
  pages = {045429},
  numpages = {14},
  year = {2011},
  month = {Jul},
  publisher = {American Physical Society},
  doi = {10.1103/PhysRevB.84.045429},
  url = {https://link.aps.org/doi/10.1103/PhysRevB.84.045429}
}

@article{NatafCiuti2010_2,
  author       = {Nataf, Pierre and Ciuti, Cristiano},
  title        = {Vacuum Degeneracy of a Circuit {QED} System in the Ultrastrong Coupling Regime},
  journal      = {Phys. Rev. Lett.},
  volume       = {104},
  number       = {2},
  pages        = {023601},
  year         = {2010},
  publisher    = {American Physical Society},
  doi          = {10.1103/PhysRevLett.104.023601}
}

@article{NatafCiuti2011a,
  author       = {Nataf, Pierre and Ciuti, Cristiano},
  title        = {Protected Quantum Computation with Multiple Resonators in Ultrastrong Coupling Circuit {QED}},
  journal      = {Phys. Rev. Lett.},
  volume       = {107},
  number       = {19},
  pages        = {190402},
  year         = {2011},
  publisher    = {American Physical Society},
  doi          = {10.1103/PhysRevLett.107.190402}
}

@article{RomanRocheLuisZueco2021,
  author       = {Rom\'an-Roche, Juan and Luis, Fernando and Zueco, David},
  title        = {Photon Condensation and Enhanced Magnetism in Cavity {QED}},
  journal      = {Phys. Rev. Lett.},
  volume       = {127},
  number       = {16},
  pages        = {167201},
  year         = {2021},
  publisher    = {American Physical Society},
  doi          = {10.1103/PhysRevLett.127.167201}
}

@article{BambaOgawa2014,
  author       = {Bamba, Motoaki and Ogawa, Tetsuo},
  title        = {Stability of polarizable materials against superradiant phase transition},
  journal      = {Phys. Rev. A},
  volume       = {90},
  number       = {6},
  pages        = {063825},
  year         = {2014},
  publisher    = {American Physical Society},
  doi          = {10.1103/PhysRevA.90.063825}
}

@article{Chirolli2012,
  author       = {Chirolli, Luca and Polini, Marco and Giovannetti, Vittorio and MacDonald, A. H. and Guinea, Francisco},
  title        = {Drude weight, cyclotron resonance, and the {{Dicke}} model of graphene cavity {QED}},
  journal      = {Phys. Rev. Lett.},
  volume       = {109},
  number       = {26},
  pages        = {267404},
  year         = {2012},
  publisher    = {American Physical Society},
  doi          = {10.1103/PhysRevLett.109.267404}
}

@article{Todorov2012,
  title = {Intersubband polaritons in the electrical dipole gauge},
  author = {Todorov, Yanko and Sirtori, Carlo},
  journal = {Phys. Rev. B},
  volume = {85},
  issue = {4},
  pages = {045304},
  numpages = {20},
  year = {2012},
  month = {Jan},
  publisher = {American Physical Society},
  doi = {10.1103/PhysRevB.85.045304},
  url = {https://link.aps.org/doi/10.1103/PhysRevB.85.045304}
}

@article{HaynEmaryBrandes2012,
  author       = {Hayn, Mathias and Emary, Clive and Brandes, Tobias},
  title        = {Superradiant phase transition in a model of three-level systems interacting with two bosonic modes},
  journal      = {Phys. Rev. A},
  volume       = {86},
  number       = {6},
  pages        = {063822},
  year         = {2012},
  publisher    = {American Physical Society},
  doi          = {10.1103/PhysRevA.86.063822}
}

@article{GawedzkiRzazewski1981,
  author       = {Gaw\k{e}dzki, K. and Rza{\.z}ewski, K.},
  title        = {No-go theorem for the superradiant phase transition without the dipole approximation},
  journal      = {Phys. Rev. A},
  volume       = {23},
  number       = {5},
  pages        = {2134--2136},
  year         = {1981},
  publisher    = {American Physical Society},
  doi          = {10.1103/PhysRevA.23.2134}
}

@article{BialynickiBirulaRzazewski1979,
  author       = {Bialynicki-Birula, I. and Rza{\.z}ewski, K.},
  title        = {No-go theorem concerning the superradiant phase transition in atomic systems},
  journal      = {Phys. Rev. A},
  volume       = {19},
  number       = {1},
  pages        = {301--303},
  year         = {1979},
  publisher    = {American Physical Society},
  doi          = {10.1103/PhysRevA.19.301}
}

@article{Andolina2019,
  author       = {Andolina, G. M. and Pellegrino, F. M. D. and Giovannetti, V. and MacDonald, A. H. and Polini, M.},
  title        = {Cavity quantum electrodynamics of strongly correlated electron systems: A no-go theorem for photon condensation},
  journal      = {Phys. Rev. B},
  volume       = {100},
  number       = {12},
  pages        = {121109(R)},
  year         = {2019},
  publisher    = {American Physical Society},
  doi          = {10.1103/PhysRevB.100.121109}
}

@article{Stokes2020,
  title = {Uniqueness of the Phase Transition in Many-Dipole Cavity Quantum Electrodynamical Systems},
  author = {Stokes, Adam and Nazir, Ahsan},
  journal = {Phys. Rev. Lett.},
  volume = {125},
  issue = {14},
  pages = {143603},
  numpages = {6},
  year = {2020},
  month = {Sep},
  publisher = {American Physical Society},
  doi = {10.1103/PhysRevLett.125.143603},
  url = {https://link.aps.org/doi/10.1103/PhysRevLett.125.143603}
}

@article{Rzazewski1975,
  author       = {Rza{\.z}ewski, K. and W{\'o}dkiewicz, K. and {\.Z}akowicz, W.},
  title        = {Phase Transitions, Two-Level Atoms, and the ${A}^{2}$ Term},
  journal      = {Phys. Rev. Lett.},
  volume       = {35},
  number       = {7},
  pages        = {432--434},
  year         = {1975},
  publisher    = {American Physical Society},
  doi          = {10.1103/PhysRevLett.35.432}
}

@article{Keeling2007,
  author       = {Keeling, Jonathan},
  title        = {Coulomb interactions, gauge invariance, and phase transitions of the {{Dicke}} model},
  journal      = {J. Condens. Matter Phys.},
  volume       = {19},
  number       = {29},
  pages        = {295213},
  year         = {2007},
  month        = jul,
  doi          = {10.1088/0953-8984/19/29/295213},
  publisher    = {IOP Publishing}
}

@article{Brennecke2013PNAS,
  author  = {Brennecke, F. and Mottl, R. and Baumann, K. and Landig, R. and Donner, T. and Esslinger, T.},
  title   = {Real-time observation of fluctuations at the driven-dissipative {Dicke} phase transition},
  journal = {Proc. Natl. Acad. Sci.},
  volume  = {110},
  number  = {29},
  pages   = {11763--11767},
  year    = {2013},
  doi     = {10.1073/pnas.1306993110}
}

@article{Carmichael1973,
title = {Higher order corrections to the {Dicke} superradiant phase transition},
journal = {Phys. Lett. A},
volume = {46},
number = {1},
pages = {47-48},
year = {1973},
issn = {0375-9601},
doi = {https://doi.org/10.1016/0375-9601(73)90679-8},
url = {https://www.sciencedirect.com/science/article/pii/0375960173906798},
author = {H.J. Carmichael and C.W. Gardiner and D.F. Walls}
}

@article{KirtonKeeling2017PRL,
  author  = {Kirton, Peter and Keeling, Jonathan},
  title   = {Suppressing and Restoring the {Dicke} Superradiance Transition by Dephasing and Decay},
  journal = {Phys. Rev. Lett.},
  volume  = {118},
  number  = {12},
  pages   = {123602},
  year    = {2017},
  doi     = {10.1103/PhysRevLett.118.123602}
}

@article{Ritsch2013RMP,
  author  = {Ritsch, Helmut and Domokos, Peter and Brennecke, Ferdinand and Esslinger, Tilman},
  title   = {Cold atoms in cavity-generated dynamical optical potentials},
  journal = {Rev. Mod. Phys.},
  volume  = {85},
  number  = {2},
  pages   = {553--601},
  year    = {2013},
  doi     = {10.1103/RevModPhys.85.553}
}

@article{Rokaj2018,
doi = {10.1088/1361-6455/aa9c99},
url = {https://doi.org/10.1088/1361-6455/aa9c99},
year = {2018},
month = {jan},
publisher = {IOP Publishing},
volume = {51},
number = {3},
pages = {034005},
author = {Rokaj, Vasil and Welakuh, Davis M and Ruggenthaler, Michael and Rubio, Angel},
title = {Light--matter interaction in the long-wavelength limit: no ground-state without dipole self-energy},
journal = {J. Phys. B:At., Mol. Opt. Phys.},
}

@article{DallaTorre2013PRA,
  author  = {Dalla Torre, E. G. and Diehl, S. and Lukin, M. D. and Sachdev, S. and Strack, P.},
  title   = {Keldysh approach for nonequilibrium phase transitions in quantum optics: Beyond the {Dicke} model in optical cavities},
  journal = {Phys. Rev. A},
  volume  = {87},
  number  = {2},
  pages   = {023831},
  year    = {2013},
  doi     = {10.1103/PhysRevA.87.023831}
}

@article{Konya2012PRA,
  author  = {K{\'o}nya, G. and Nagy, D. and Szirmai, G. and Domokos, P.},
  title   = {Finite-size scaling in the quantum phase transition of the open-system {Dicke} model},
  journal = {Phys. Rev. A},
  volume  = {86},
  number  = {1},
  pages   = {013641},
  year    = {2012},
  doi     = {10.1103/PhysRevA.86.013641}
}

@article{Klinder2015PNAS,
  author  = {Klinder, J. and Ke\ss{}ler, H. and Wolke, M. and Mathey, L. and Hemmerich, A.},
  title   = {Dynamical phase transition in the open {Dicke} model},
  journal = {Proc. Natl. Acad. Sci.},
  volume  = {112},
  number  = {11},
  pages   = {3290--3295},
  year    = {2015},
  doi     = {10.1073/pnas.1417132112}
}

@article{Baden2014PRL,
  author  = {Baden, M. P. and Arnold, K. J. and Grimsmo, A. L. and Parkins, S. and Barrett, M. D.},
  title   = {Realization of the {Dicke} Model Using Cavity-Assisted Raman Transitions},
  journal = {Phys. Rev. Lett.},
  volume  = {113},
  number  = {2},
  pages   = {020408},
  year    = {2014},
  doi     = {10.1103/PhysRevLett.113.020408}
}

@article{Dimer2007,
  title = {Proposed realization of the {Dicke}-model quantum phase transition in an optical cavity {QED} system},
  author = {Dimer, F. and Estienne, B. and Parkins, A. S. and Carmichael, H. J.},
  journal = {Phys. Rev. A},
  volume = {75},
  issue = {1},
  pages = {013804},
  numpages = {14},
  year = {2007},
  month = {Jan},
  publisher = {American Physical Society},
  doi = {10.1103/PhysRevA.75.013804},
  url = {https://link.aps.org/doi/10.1103/PhysRevA.75.013804}
}

@article{WangHioe1973,
  author       = {Wang, Y. K. and Hioe, F. T.},
  title        = {Phase Transition in the {Dicke} Model of Superradiance},
  journal      = {Phys. Rev. A},
  volume       = {7},
  number       = {3},
  pages        = {831--836},
  year         = {1973},
  publisher    = {American Physical Society},
  doi          = {10.1103/PhysRevA.7.831}
}

@article{Zhang2016Collective,
  author    = {Qi Zhang and Minhan Lou and Xinwei Li and John L. Reno and Wei Pan and John D. Watson and Michael J. Manfra and Junichiro Kono},
  title     = {Collective non-perturbative coupling of 2D electrons with high-quality-factor terahertz cavity photons},
  journal   = {Nature Physics},
  volume    = {12},
  pages     = {1005--1011},
  year      = {2016},
  doi       = {10.1038/nphys3850},
  url       = {https://doi.org/10.1038/nphys3850}
}

@article{Bayer2017,
author = {Bayer, Andreas and Pozimski, Marcel and Schambeck, Simon and Schuh, Dieter and Huber, Rupert and Bougeard, Dominique and Lange, Christoph},
title = {Terahertz Light--Matter Interaction beyond Unity Coupling Strength},
journal = {Nano Letters},
volume = {17},
number = {10},
pages = {6340-6344},
year = {2017},
doi = {10.1021/acs.nanolett.7b03103},
URL = {https://doi.org/10.1021/acs.nanolett.7b03103},
}

@article{Baumann2011Dicke,
  title = {Exploring Symmetry Breaking at the Dicke Quantum Phase Transition},
  author = {Baumann, K. and Mottl, R. and Brennecke, F. and Esslinger, T.},
  journal = {Phys. Rev. Lett.},
  volume = {107},
  issue = {14},
  pages = {140402},
  numpages = {5},
  year = {2011},
  month = {Sep},
  publisher = {American Physical Society},
  doi = {10.1103/PhysRevLett.107.140402},
  url = {https://link.aps.org/doi/10.1103/PhysRevLett.107.140402}
}

@article{Dicke1954,
  title = {Coherence in Spontaneous Radiation Processes},
  author = {Dicke, R. H.},
  journal = {Phys. Rev.},
  volume = {93},
  issue = {1},
  pages = {99--110},
  numpages = {0},
  year = {1954},
  month = {Jan},
  publisher = {American Physical Society},
  doi = {10.1103/PhysRev.93.99},
  url = {https://link.aps.org/doi/10.1103/PhysRev.93.99}
}

@article{Scalari2012,
author = {G. Scalari  and C. Maissen  and D. Tur\v{c}inkov\'{a}  and D. Hagenm\"{u}ller  and S. De Liberato  and C. Ciuti  and C. Reichl  and D. Schuh  and W. Wegscheider  and M. Beck  and J. Faist },
title = {Ultrastrong Coupling of the Cyclotron Transition of a 2D Electron Gas to a THz Metamaterial},
journal = {Science},
volume = {335},
number = {6074},
pages = {1323-1326},
year = {2012},
doi = {10.1126/science.1216022},
URL = {https://www.science.org/doi/abs/10.1126/science.1216022},
}

@article{Li2018Observation,
  author       = {Xinwei Li and Motoaki Bamba and Ning Yuan and Qi Zhang and
                  Yage Zhao and Maolin Xiang and Kai Xu and Zuanming Jin and
                  Wei Ren and Guohong Ma and Shixun Cao and Dmitry Turchinovich and
                  Junichiro Kono},
  title        = {Observation of {Dicke} cooperativity in magnetic interactions},
  journal      = {Science},
  volume       = {361},
  number       = {6404},
  pages        = {794--797},
  year         = {2018},
  doi          = {10.1126/science.aat5162},
  url          = {https://doi.org/10.1126/science.aat5162}
}

@article{Nagy2010,
  title = {{Dicke}-Model Phase Transition in the Quantum Motion of a Bose-Einstein Condensate in an Optical Cavity},
  author = {Nagy, D. and K\'onya, G. and Szirmai, G. and Domokos, P.},
  journal = {Phys. Rev. Lett.},
  volume = {104},
  issue = {13},
  pages = {130401},
  numpages = {4},
  year = {2010},
  month = {Apr},
  publisher = {American Physical Society},
  doi = {10.1103/PhysRevLett.104.130401},
  url = {https://link.aps.org/doi/10.1103/PhysRevLett.104.130401}
}

@article{Liu2023,
  title = {Switchable superradiant phase transition with Kerr magnons},
  author = {Liu, Gang and Xiong, Wei and Ying, Zu-Jian},
  journal = {Phys. Rev. A},
  volume = {108},
  issue = {3},
  pages = {033704},
  numpages = {7},
  year = {2023},
  month = {Sep},
  publisher = {American Physical Society},
  doi = {10.1103/PhysRevA.108.033704},
  url = {https://link.aps.org/doi/10.1103/PhysRevA.108.033704}
}

@article{KimEtAl2024,
  author       = {Kim, Dasom and Dasgupta, Sohail and Ma, Xiaoxuan and Park, Joong-Mok and Wei, Hao-Tian and Luo, Liang and Doumani, Jacques and Li, Xinwei and Yang, Wanting and Cheng, Di and Kim, Richard H.J. and Everitt, Henry O. and Kimura, Shojiro and Nojiri, Hiroyuki and Wang, Jigang and Cao, Shixun and Bamba, Motoaki and Hazzard, Kaden R.A. and Kono, Junichiro},
  title        = {Observation of the Magnonic {Dicke} Superradiant Phase Transition},
  journal      = {arXiv preprint},
  archivePrefix= {arXiv},
  eprint       = {2401.01873},
  year         = {2024}
}

@article{MarquezPeracaEtAl2024,
  author       = {Marquez Peraca, Nicolas and others and Kono, Junichiro},
  title        = {Quantum simulation of an extended {Dicke} model with a magnetic solid},
  journal      = {Commun. Mater.},
  volume       = {10},
  pages        = {79},
  year         = {2024}
}

@article{BambaLiMarquezPeracaKono2022,
  author       = {Bamba, Motoaki and Li, Xinwei and Marquez Peraca, Nicolas and Kono, Junichiro},
  title        = {Magnonic superradiant phase transition},
  journal      = {Commun. Phys.},
  volume       = {5},
  pages        = {3},
  year         = {2022},
  publisher    = {Nature Publishing Group},
  doi          = {10.1038/s42005-021-00785-z}
}

@article{HeppLieb1973,
  author       = {Hepp, K. and Lieb, E. H.},
  title        = {On the Superradiant Phase Transition for Molecules in a Quantized Radiation Field: The {Dicke} Maser Model},
  journal      = {Ann. Phys.},
  volume       = {76},
  number       = {2},
  pages        = {360--404},
  year         = {1973},
  publisher    = {Academic Press},
  doi          = {10.1016/0003-4916(73)90039-0}
}

@article{EmaryBrandes2003,
  author       = {Emary, Clive and Brandes, Tobias},
  title        = {Chaos and the Quantum Phase Transition in the {Dicke} Model},
  journal      = {Phys. Rev. E},
  volume       = {67},
  number       = {6},
  pages        = {066203},
  year         = {2003},
  publisher    = {American Physical Society},
  doi          = {10.1103/PhysRevE.67.066203}
}

@article{Ciuti2005,
  title = {Quantum vacuum properties of the intersubband cavity polariton field},
  author = {Ciuti, C. and Bastard, G. and Carusotto, I.},
  journal = {Phys. Rev. B},
  volume = {72},
  issue = {11},
  pages = {115303},
  numpages = {9},
  year = {2005},
  month = {Sep},
  publisher = {American Physical Society},
  doi = {10.1103/PhysRevB.72.115303},
  url = {https://link.aps.org/doi/10.1103/PhysRevB.72.115303}
}

@article{Bloch2022,
  title = {Strongly correlated electron--photon systems},
  author = {Bloch, Jacqueline and Cavalleri, Andrea and Galitski, Victor and Hafezi, Mohammad and Rubio, Angel},
  journal = {Nature},
  volume = {606},
  number = {7912},
  pages = {41--48},
  year = {2022},
  doi = {10.1038/s41586-022-04726-w},
  url = {https://doi.org/10.1038/s41586-022-04726-w}
}

@article{Hubener2024,
  title = {Quantum materials engineering by structured cavity vacuum fluctuations},
  author = {H{\"u}bener, Hannes and Vi{\~n}as Bostr{\"o}m, Emil and Claassen, Martin and Latini, Simone and Rubio, Angel},
  journal = {Mater. Quantum Technol.},
  volume = {4},
  number = {2},
  pages = {023002},
  year = {2024},
  doi = {10.1088/2633-4356/ad4e8b},
  url = {https://doi.org/10.1088/2633-4356/ad4e8b}
}

@article{PhysRevLett.134.063602,
  title = {Floquet Engineering the Quantum Rabi Model in the Ultrastrong Coupling Regime},
  author = {Akbari, Kamran and Nori, Franco and Hughes, Stephen},
  journal = {Phys. Rev. Lett.},
  volume = {134},
  issue = {6},
  pages = {063602},
  numpages = {8},
  year = {2025},
  month = {Feb},
  publisher = {American Physical Society},
  doi = {10.1103/PhysRevLett.134.063602},
  url = {https://link.aps.org/doi/10.1103/PhysRevLett.134.063602}
}

@article{PhysRevResearch.7.013271,
  title = {Quantum phase transitions in many-dipole light-matter systems},
  author = {Lamberto, Daniele and Di Stefano, Omar and Hughes, Stephen and Nori, Franco and Savasta, Salvatore},
  journal = {Phys. Rev. Res.},
  volume = {7},
  issue = {1},
  pages = {013271},
  numpages = {16},
  year = {2025},
  month = {Mar},
  publisher = {American Physical Society},
  doi = {10.1103/PhysRevResearch.7.013271},
  url = {https://link.aps.org/doi/10.1103/PhysRevResearch.7.013271}
}

@article{Garcia-Vidal2021,
author = {Francisco J. Garcia-Vidal  and Cristiano Ciuti  and Thomas W. Ebbesen },
title = {Manipulating matter by strong coupling to vacuum fields},
journal = {Science},
volume = {373},
number = {6551},
pages = {eabd0336},
year = {2021},
doi = {10.1126/science.abd0336},
URL = {https://www.science.org/doi/abs/10.1126/science.abd0336},
}

@article{Schlawin2022,
    author = {Schlawin, F. and Kennes, D. M. and Sentef, M. A.},
    title = {Cavity quantum materials},
    journal = {Appl. Phys. Rev.},
    volume = {9},
    number = {1},
    pages = {011312},
    year = {2022},
    month = {02},
    issn = {1931-9401},
    doi = {10.1063/5.0083825},
    url = {https://doi.org/10.1063/5.0083825},
}

@article{PhysRevB.81.235303,
  title = {Ultrastrong coupling between a cavity resonator and the cyclotron transition of a two-dimensional electron gas in the case of an integer filling factor},
  author = {Hagenm\"uller, David and De Liberato, Simone and Ciuti, Cristiano},
  journal = {Phys. Rev. B},
  volume = {81},
  issue = {23},
  pages = {235303},
  numpages = {9},
  year = {2010},
  month = {Jun},
  publisher = {American Physical Society},
  doi = {10.1103/PhysRevB.81.235303},
  url = {https://link.aps.org/doi/10.1103/PhysRevB.81.235303}
}

@article{Eckardt_2015,
doi = {10.1088/1367-2630/17/9/093039},
url = {https://doi.org/10.1088/1367-2630/17/9/093039},
year = {2015},
month = {sep},
publisher = {IOP Publishing},
volume = {17},
number = {9},
pages = {093039},
author = {Eckardt, André and Anisimovas, Egidijus},
title = {High-frequency approximation for periodically driven quantum systems from a Floquet-space perspective},
journal = {New J. Phys.},
}

@article{HagenmullerDavid2016AdCe,
  title = {All-optical dynamical Casimir effect in a three-dimensional terahertz photonic band gap},
  author = {Hagenm\"uller, David},
  journal = {Phys. Rev. B},
  volume = {93},
  issue = {23},
  pages = {235309},
  numpages = {17},
  year = {2016},
  month = {Jun},
  publisher = {American Physical Society},
  doi = {10.1103/PhysRevB.93.235309},
  url = {https://link.aps.org/doi/10.1103/PhysRevB.93.235309}
}

@article{DeLiberato2007,
  title = {Quantum Vacuum Radiation Spectra from a Semiconductor Microcavity with a Time-Modulated Vacuum Rabi Frequency},
  author = {Liberato, Simone De and Ciuti, Cristiano and Carusotto, Iacopo},
  journal = {Phys. Rev. Lett.},
  volume = {98},
  issue = {10},
  pages = {103602},
  numpages = {4},
  year = {2007},
  month = {Mar},
  publisher = {American Physical Society},
  doi = {10.1103/PhysRevLett.98.103602},
  url = {https://link.aps.org/doi/10.1103/PhysRevLett.98.103602}
}

@article{ZhangZhenya2025SsiS,
author = {Zhang, Zhenya and Kanega, Minoru and Maruyama, Kei and Kurihara, Takayuki and Nakajima, Makoto and Tachizaki, Takehiro and Sato, Masahiro and Kanemitsu, Yoshihiko and Hirori, Hideki},
address = {London},
issn = {1476-1122},
journal = {Nat. Mater},
number = {2},
pages = {219-225},
publisher = {Nature Publishing Group UK},
title = {Spin switching in {Sm$_{0.7}$Er$_{0.3}$FeO$_{3}$} triggered by terahertz magnetic-field pulses},
url = {https://doi.org/10.1038/s41563-024-02034-4},
doi = {10.1038/s41563-024-02034-4},
volume = {24},
year = {2025},
}

@article{ZybertM.2017Llas,
  title = {Landau levels and shallow donor states in {GaAs/AlGaAs} multiple quantum wells at megagauss magnetic fields},
  author = {Zybert, M. and Marchewka, M. and Sheregii, E. M. and Rickel, D. G. and Betts, J. B. and Balakirev, F. F. and Gordon, M. and Stier, A. V. and Mielke, C. H. and Pfeffer, P. and Zawadzki, W.},
  journal = {Phys. Rev. B},
  volume = {95},
  issue = {11},
  pages = {115432},
  numpages = {7},
  year = {2017},
  month = {Mar},
  publisher = {American Physical Society},
  doi = {10.1103/PhysRevB.95.115432},
  url = {https://link.aps.org/doi/10.1103/PhysRevB.95.115432}
}

@article{PFEFFERP1990CeiG,
  title = {Conduction electrons in {GaAs}: Five-level k\ensuremath{\cdot}p theory and polaron effects},
  author = {Pfeffer, P. and Zawadzki, W.},
  journal = {Phys. Rev. B},
  volume = {41},
  issue = {3},
  pages = {1561--1576},
  numpages = {0},
  year = {1990},
  month = {Jan},
  publisher = {American Physical Society},
  doi = {10.1103/PhysRevB.41.1561},
  url = {https://link.aps.org/doi/10.1103/PhysRevB.41.1561}
}
\end{document}


\begin{CJK*}{UTF8}{bsmi}

\title{Supplemental Material for\\ Floquet Engineering of a Quasiequilibrium Superradiant Phase Transition in Landau Polaritons}

\author{Wen-Hua Wu (吳文華)}
\affiliation{Applied Physics Graduate Program, Smalley-Curl Institute, Rice University, Houston, Texas 77005, USA}
\affiliation{Department of Electrical and Computer Engineering, Rice University, Houston, Texas 77005, USA}

\author{Fuyang Tay}
\affiliation{Department of Physics, Columbia University, New York, NY 10027, USA}
\affiliation{Department of Chemistry, Columbia University, New York, NY 10027, USA}

\author{Mengqian Che}
\affiliation{Department of Electrical and Computer Engineering, Rice University, Houston, Texas 77005, USA}

\author{Andrey Baydin}
\affiliation{Department of Electrical and Computer Engineering, Rice University, Houston, Texas 77005, USA}
\affiliation{Smalley-Curl Institute, Rice University, Houston, Texas 77005, USA}
\affiliation{Rice Advanced Materials Institute, Rice University, Houston, Texas 77005, USA}

\author{Junichiro Kono}
\affiliation{Department of Electrical and Computer Engineering, Rice University, Houston, Texas 77005, USA}
\affiliation{Smalley-Curl Institute, Rice University, Houston, Texas 77005, USA}
\affiliation{Rice Advanced Materials Institute, Rice University, Houston, Texas 77005, USA}
\affiliation{Department of Physics and Astronomy, Rice University, Houston, Texas 77005, USA}
\affiliation{Department of Materials Science and NanoEngineering, Rice University, Houston, Texas 77005, USA}

\author{David Hagenm\"{u}ller}
\email{david.hagenmuller@ipcms.unistra.fr}
\affiliation{IPCMS (UMR 7504), Université de Strasbourg and CNRS, Strasbourg, France}

\begin{abstract}
In this Supplemental Material, we present details of the theoretical framework and numerical methods. In Sec.~\ref{sec:Hamilt_eq}, we derive the equilibrium Landau polariton Hamiltonian in second quantization. Section~\ref{Floquet_exp} analyzes the dynamics of the time-dependent Floquet Hamiltonian in the normal phase. In Sec.~\ref{cutoff_arti}, we demonstrate that an infinite ladder of equally spaced Landau levels does not support a SRPT and that imposing a rigid cutoff leads to unphysical results. Section~\ref{breakdown_2L} shows the breakdown of the two-level approximation in the superradiant phase. Finally, Secs.~\ref{numerical_details} and \ref{analytics_order} provide details of the numerical procedure and a perturbative calculation of the expectation values in the presence of band nonparabolicity.
\end{abstract}

\maketitle
\end{CJK*}

\tableofcontents

\section{Hamiltonian at equilibrium}
\label{sec:Hamilt_eq}

In this section, we derive the microscopic Hamiltonian in its second-quantized form given by Eq.~(1) of the main text. At equilibrium, the Hamiltonian reads $H=\hat{H}_{\textrm{cav}}+\sum_{k} \mathcal{H}_{k}$, where $\hat{H}_{\textrm{cav}} = \hbar \omega_{\textrm{cav}} \hat{a}^{\dagger} \hat{a}$ is the bare cavity photon Hamiltonian, and
\begin{align}
\mathcal{H}_{k} &=\frac{1}{2m} \left[\mathbf{p} +e \mathbf{A}(\mathbf{r})\right]^2 =\mathcal{H}_{0,k} + \mathcal{V}_{k} + \mathcal{H}_{\textrm{dia}}
\end{align}
is the single-electron Hamiltonian. Here $\mathbf{p}=p_{x} \mathbf{u}_{x} + p_{y} \mathbf{u}_{y}$ is the electron momentum, and $\mathbf{A}(\mathbf{r}) =\mathbf{A}_{0}(\mathbf{r}) + \hat{\mathbf{A}}_{\textrm{cav}}(\mathbf{r})$ is the total vector potential, which consists of a static component $\mathbf{A}_{0}(\mathbf{r})=Bx \mathbf{u}_{y}$ and the cavity mode contribution $$\hat{\mathbf{A}}_{\textrm{cav}}(\mathbf{r})=\sqrt{\frac{\hbar}{\epsilon_{0} \omega_{\textrm{cav}}S L}} (\hat{a}+\hat{a}^{\dagger})\mathbf{u}_{x}.$$ The single-electron Hamiltonian $\mathcal{H}_{k}$ is decomposed into three contributions:
\begin{align}
\mathcal{H}_{0,k} &= \frac{1}{2m} \left[\mathbf{p} +e \mathbf{A}_{0}(\mathbf{r})\right]^2 \nonumber \\
\mathcal{V}_{k}& =\frac{e}{m} \mathbf{p} \cdot \mathbf{A}_{\textrm{cav}}(\mathbf{r}) \nonumber \\ 
\mathcal{H}_{\textrm{dia}}& = \frac{e^2}{2 m} \mathbf{A}^2_{\textrm{cav}} (\mathbf{r}).
\end{align}
Introducing the harmonic oscillator ladder operators $d_{k}$ and $d^{\dagger}_{k}$, defined by 
\begin{align}
d_{k}&=\frac{l_{0}}{\hbar \sqrt{2}} \left(p_{y} + e B x + i p_{x} \right) \nonumber \\
d^{\dagger}_{k}&=\frac{l_{0}}{\hbar \sqrt{2}} \left(p_{y} + e B x - i p_{x} \right),
\end{align}
and acting on the single particle states as $d_{k} \vert n,k \rangle =\sqrt{n} \vert n-1,k \rangle$ and $d^{\dagger}_{k} \vert n,k \rangle =\sqrt{n+1} \vert n+1,k \rangle$, 
the single-electron Hamiltonian can be written, up to a constant term $\hbar \omega_{0}/2$ and at resonance, $\omega_{\textrm{cav}}=\omega_{0}$, as $\mathcal{H}_{k}=\mathcal{H}_{0,k} + \mathcal{V}_{k}  + \mathcal{H}_{\textrm{dia}}$, where
\begin{align}
\mathcal{H}_{0,k} & = \hbar\omega_0 d^\dagger_{k} d_{k} \nonumber \\
\mathcal{V}_{k} &= \frac{\hbar g_{0}}{\sqrt{N}} \left(d_{k} + d^\dagger_{k}\right) \left(\hat{a} + \hat{a}^{\dagger}\right) \nonumber \\
\mathcal{H}_{\textrm{dia}} & = \frac{\hbar g^2_{0}}{N\omega_{0}} \left(\hat{a} + \hat{a}^{\dagger}\right)^2.
\label{eq:eq_distant_LL_ham1}
\end{align}
Using the field operators $\hat{\Psi} (\mathbf{r})=\sum_{n,k} \langle \mathbf{r} \vert n,k \rangle \hat{c}_{n,k}$, where $\hat{c}_{n,k}$ annihilates an electron in the single-particle state $\vert n,k\rangle$, and its hermitian conjugate $\hat{\Psi}^{\dagger} (\mathbf{r})$, it can be checked that the second-quantized Hamiltonian $$\hat{H} = \hat{H}_{\textrm{cav}} + \sum_{k} \int \! d\mathbf{r} \, \hat{\Psi}^{\dagger} (\mathbf{r}) \mathcal{H}_{k} \hat{\Psi} (\mathbf{r})$$ coincides with Eq.~(1) of the main text. The $\hat{\mathbf{A}}^2_{\textrm{cav}}$-term is derived using 
\begin{equation}
\int \! d\mathbf{r} \; \hat{\Psi}^{\dagger} (\mathbf{r}) \hat{\Psi} (\mathbf{r}) \rightarrow  \sum_{n,k} \langle G^{(0)}\vert \hat{c}^{\dagger}_{n,k} \hat{c}_{n,k} \vert G^{(0)} \rangle = \rho S= N \nu,
\end{equation} 
where $\vert G^{(0)} \rangle$ denotes the ground state of the Landau polariton system in the absence of light-matter coupling. Within the reduced Hilbert space consisting of the two LLs adjacent to the Fermi energy, namely the highest occupied level $n=\nu-1$ and the lowest unoccupied level $n=\nu$, the Hamiltonian Eq.~(1) of the main text can be rewritten as 
\begin{align}
\label{eq:hopfield_model_equilibrium}
\hat{H} = \hbar\omega_{0} \hat{b}^\dagger \hat{b} 
+ \hbar\omega_{0} \hat{a}^\dagger \hat{a}
+ \hbar D \left(\hat{a}^\dagger + \hat{a}\right)^2  + \hbar \Omega_{0} \left(\hat{a}^\dagger + \hat{a}\right) \left(\hat{b}^\dagger + \hat{b}\right),
\end{align}
where the collective cyclotron resonance (CR) annihilation operator reads $\hat{b} = \frac{1}{\sqrt{N}} \sum_{k} 
\hat{c}^{\dagger}_{\nu-1,k}\hat{c}^{\vphantom{\dagger}}_{\nu,k}$.

\section{Floquet hamiltonian dynamics}
\label{Floquet_exp} 

In this section, we outline the computation of the dynamics generated by the Floquet Hamiltonian (at resonance $\omega_{\textrm{cav}}=\omega_{0}$)
\begin{align}
\label{eq:hopfield_model}
\hat{H}(t) = \hbar\omega (t) \hat{b}^\dagger \hat{b} 
+ \hbar\omega_{0} \hat{a}^\dagger \hat{a}
+ \hbar D \left(\hat{a}^\dagger + \hat{a}\right)^2  + \hbar \Omega (t) \left(\hat{a}^\dagger + \hat{a}\right)
\left(\hat{b}^\dagger + \hat{b}\right),
\end{align}
where $B(t) = B_0 + \Delta B \cos(\omega_p t)$, $\omega (t) = \omega_0 \left[ 1 + \varepsilon \cos(\omega_p t) \right]$, and $\Omega (t)=g(t)\sqrt{\nu}$, with $g(t) = g_0 \sqrt{\left| 1 + \varepsilon \cos(\omega_p t) \right|}$.
The evolution of the system is governed by the equation of motion 
\begin{align}
\dot{\Phi}(t) = A(t)\Phi(t) + \Phi(t)A(t),
\label{eq_of_motio}
\end{align}
with $\Phi \equiv \hat{\mathbf{v}}\hat{\mathbf{v}}^{\dagger}$ and $\hat{\mathbf{v}}\equiv (\hat{a}\,\hat{b}\,\hat{a}^\dagger\,\hat{b}^{\dagger})^{\textrm{T}}$ is a column vector. The solutions of the equation of motion satisfy $\Phi(T) = M\Phi(0)$, where the so-called monodromy matrix $M$ (or Floquet operator) is defined as the evolution
operator over the period $T$. The eigenvalues of the Floquet operator can be written in the form $e^{\mu_{j} T}$, where $\mu_{j}$ are the Floquet exponents. Small $\mathrm{Re} \{\mu_{j}\}$ indicate that the system practically does not evolve over the period $T$ while positive $\mathrm{Re} \{\mu_{j}\}$ indicate the presence of an instability characterized by exponential growth of the fields. The time-dependent matrix $A(t)$ which characterizes the dynamics reads
\begin{equation}
A(t) =-i
\begin{bmatrix}
\omega_{0} + 2D & \Omega(t) & 2D & \Omega(t) \\
\Omega(t) & \omega(t) & \Omega(t) & 0 \\
-2D & -\Omega(t) & -\omega_{0} - 2D & -\Omega(t) \\
-\Omega(t) & 0 & -\Omega(t) & -\omega(t)
\end{bmatrix}.
\end{equation}
Figure~1(c) of the main text shows the maximum real part of the Floquet exponents, $\max_j \mathrm{Re}\{\mu_j\}$, as a function of the AC magnetic-field amplitude $\Delta B$ and equilibrium collective coupling strength $\Omega_0=g_0\sqrt{\nu}$ for $\omega_{\mathrm p}/2\pi=20~\mathrm{THz}$. Positive values identify a Floquet-instability region above a critical threshold.

\begin{figure}
    \centering
    \begin{minipage}[t]{0.5\textwidth}
    \vspace{0pt}
        \includegraphics[width=\linewidth]{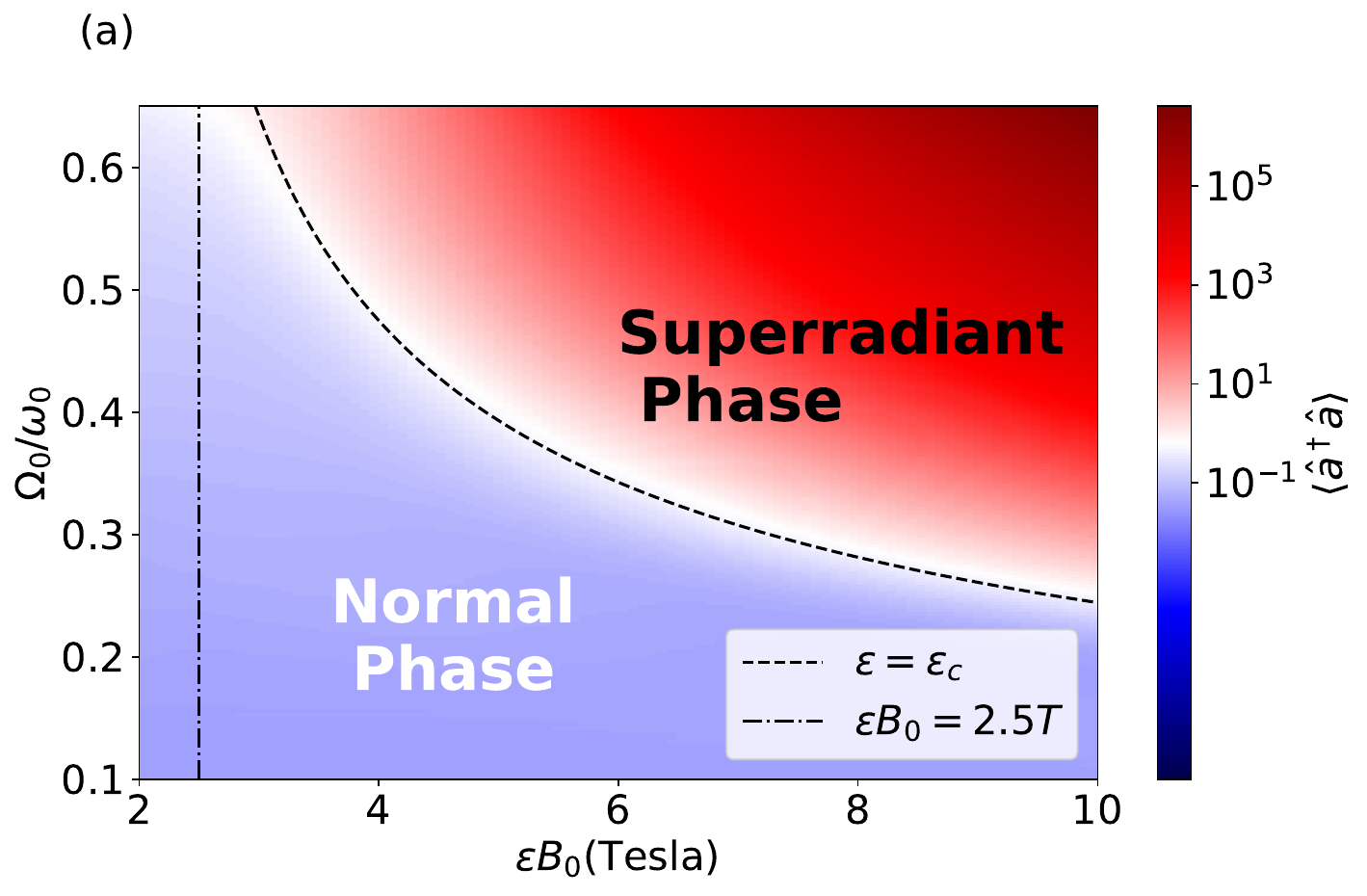}
    \end{minipage}
    \includegraphics[width=0.6\linewidth]{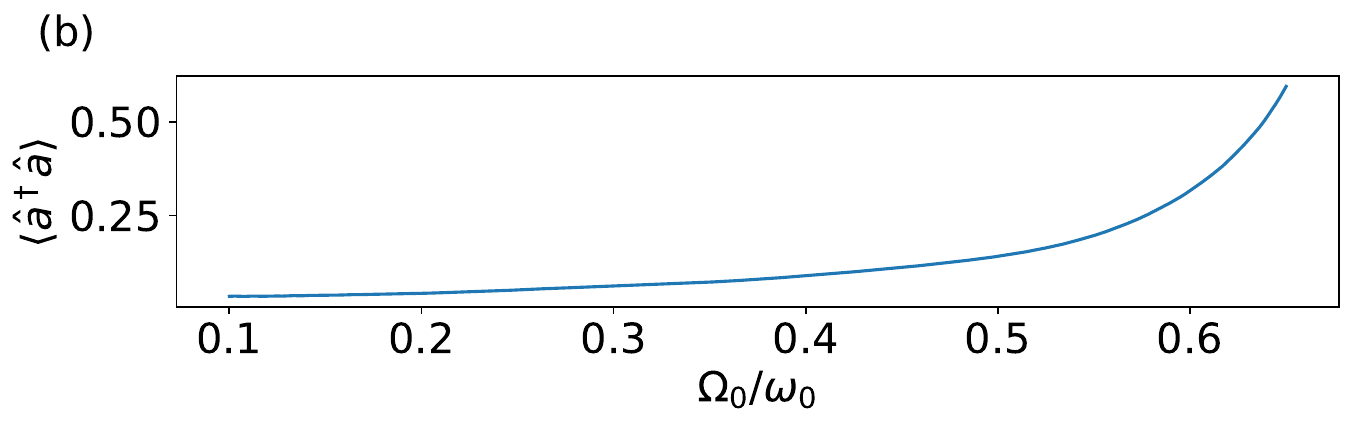}
    \caption{(a)~Time-averaged intracavity photon number $\langle \hat{a}^\dagger \hat{a} \rangle$ as a function of the equilibrium coupling strength $\Omega_0$ and the pulse amplitude $\varepsilon B_0$ for a finite-duration modulation. The black dashed curve marks the phase boundary $\eta^2 \varepsilon_c = 1 + \omega_0^2/(4\Omega_0^2)$ obtained from the effective Floquet Hamiltonian [Eq.~(5) in the main text]. (b)~Vertical line cut at $\varepsilon B_0 = 2.5$~T [black dash-dotted line in (a)]. Parameters: $t_0 = 5$~ps, $t_\mathrm{p} = 1$~ps, $\omega_\mathrm{p}/2\pi = 20$~THz, and $\omega_0/2\pi = 0.5$~THz.}
    \label{fig:N_ph}
\end{figure}

For a finite-duration magnetic-field modulation,
\begin{equation}
\Delta B(t)=\varepsilon B_0\, e^{-(t-t_0)^2/2t_\mathrm{p}^2}\cos[\omega_\mathrm{p}(t-t_0)],
\end{equation}
we compute the time evolution of the intracavity photon number by solving Eq.~\eqref{eq_of_motio}. Figure~\ref{fig:N_ph}(a) shows the resulting dynamics as a function of the equilibrium coupling strength and relative modulation (maximal) amplitude. Once the drive is switched on, the cavity field exhibits Rabi oscillations together with a high-frequency component at twice the lower-polariton frequency [Fig.~\ref{fig:N_ph}(b)]. The latter reflects the superradiant instability through the softening of the lower-polariton mode, as discussed in the main text.

To characterize the transition, we evaluate the photon population averaged over several Rabi periods. The resulting time-averaged photon number is shown in Fig.~\ref{fig:N_ph}(c) as a function of the initial collective coupling $\Omega_0/\omega_0$. A sharp increase is observed upon crossing the dashed line, corresponding to the critical point predicted by the effective Floquet Hamiltonian [Eq.~(5) of the main text] when only the leading-order DC contribution is retained. This behavior signals the onset of the SRPT. A weaker increase of the photon population with $\Omega_0$ is also present in the normal phase. It originates from the finite equilibrium ground-state photon population, which grows with $\Omega_0$ and constitutes a hallmark of the ultrastrong-coupling regime~\cite{Ciuti2005}.

\section{Order parameter divergence and rigid-cutoff artifacts in equally-spaced LLs}
\label{cutoff_arti}

In this section, we show that Floquet-driven Landau polaritons with equally spaced LLs do not exhibit a SRPT, and that introducing a rigid cutoff in the LL index leads to cutoff-dependent order parameters. As established in Sec.~\ref{sec:Hamilt_eq}, the single-particle Hamiltonian is given by Eq.~\eqref{eq:eq_distant_LL_ham1}. Under off-resonant Floquet driving, the effect of the drive is to renormalize the paramagnetic coupling according to $\Omega_0 \to \Omega_{\mathrm{eff}}$, as discussed in the main text. 

We then adopt a mean-field approach, replacing the photonic operators $\hat{a}$ and $\hat{a}^\dagger$ by a real macroscopic field $\alpha$. The resulting total Hamiltonian takes the form $\overline{H^{\mathrm{F}}}=\hbar\omega_0 \alpha^2+ \sum_{k}\overline{\mathcal{H}_{k}^{\mathrm{F}}}$, where the single-electron Hamiltonian reads
\begin{align}
\overline{\mathcal{H}_{k}^{\mathrm{F}}} &=  \hbar \omega_{0} d^{\dagger}_{k}d_{k}  + \frac{2 \alpha\hbar \Omega_{\mathrm{eff}}}{\sqrt{N\nu}} \left(d_{k} + d^\dagger_{k} \right) + \frac{4\hbar D}{N \nu}\alpha^2 \nonumber \\
&= \hbar\omega_0
\left(d^\dagger_{k} + \frac{2\alpha \Omega_{\mathrm{eff}}}{\omega_0\sqrt{N\nu}}
\right)
\left(d_{k} + \frac{2\alpha \Omega_{\mathrm{eff}}}{\omega_0\sqrt{N\nu}}
\right)
- \frac{4 \hbar \Omega^2_{\mathrm{eff}}}{\omega_0 N\nu} \alpha^2  + \frac{4\hbar D}{N \nu}\alpha^2  \nonumber \\
&= \hbar\omega_0 \tilde{d}^{\dagger}_{k} \tilde{d}_{k} + \hbar\left(\frac{4D}{N \nu} - \frac{4 \Omega^2_{\mathrm{eff}}}{\omega_0 N\nu}\right)\alpha^2 ,
\end{align}
where
\[
\tilde{d}_{k} \equiv d_{k} + \frac{2\alpha \Omega_{\mathrm{eff}}}{\omega_0\sqrt{N\nu}},
\]
and its Hermitian conjugate define new ladder operators satisfying $\tilde{d}_{k}\ket{0}_{k}=0$, with $\ket{0}_{k}$ the $(n=0,k)$ single-particle state. The paramagnetic coupling thus induces a uniform shift of the entire LL ladder without modifying the level spacing. Using $\sum_{n,k} \langle G^{(0)}| \hat{c}^{\dagger}_{n,k} \hat{c}_{n,k} |G^{(0)} \rangle = N \nu$, the second-quantized Hamiltonian becomes
\begin{align}
\hat{\overline{H^{\mathrm{F}}}}
&= \hbar\omega_0 \alpha^2 + \sum_{k} \int \! d\mathbf{r} \, \hat{\Psi}^{\dagger} (\mathbf{r}) \overline{\mathcal{H}_{k}^{\mathrm{F}}} \hat{\Psi} (\mathbf{r}) \\
&= \hbar\left(\omega_0 + 4D - \frac{4 \Omega^2_{\mathrm{eff}}}{\omega_0}\right)\alpha^2 
+ \hbar\omega_0 \sum_{n,k} n\, \hat{c}_{n,k}^\dagger \hat{c}_{n,k}.
\label{H_F_infi}
\end{align}
The corresponding ground-state energy is therefore
\begin{align}
\label{Ground_state_infinite}
E_{G}(\alpha) 
&= \hbar\left(\omega_0 + 4D - \frac{4\Omega_{\mathrm{eff}}^2}{\omega_0}\right)\alpha^2 + E_0,
\end{align}
with $E_0 \equiv N\sum_{n=0}^{\nu-1} n\hbar\omega_0 = N \nu (\nu-1)/2$, which is purely quadratic in $\alpha$, with no higher-order contributions. As a result, $E_G(\alpha)$ does not develop a local minimum, and therefore no SRPT can occur for an infinite ladder of equally spaced LLs.

One might attempt to regularize the problem by introducing a finite cutoff in the LL index, for instance by restricting the summation in Eq.~\eqref{H_F_infi} to $n \le \Lambda$. However, imposing such a rigid cutoff in a system with equally spaced LLs inevitably introduces unphysical artifacts. Indeed, for an infinite ladder of equally spaced LLs, we have shown that the ground-state energy is strictly quadratic in $\alpha$ [see Eq.~\eqref{Ground_state_infinite}]. As a consequence, the perturbative expansion of the single-particle energy spectrum associated with the Hamiltonian $\mathcal{H}_{0,k}+\overline{\mathcal{V}_{k}^{\textrm{F}}}$, where
\begin{align}
\overline{\mathcal{V}_{k}^{\textrm{F}}}=\frac{2 \alpha\hbar \Omega_{\mathrm{eff}}}{\sqrt{N\nu}} \left(d_{k} + d^\dagger_{k}\right)
\end{align}
terminates at second order in $\alpha$. In particular, the second-order correction to the energy of the $n$-th LL reads
\begin{equation}
\mathcal{E}_n^{(2)} 
= \sum_{m \neq n} \frac{|\langle m,k|\overline{\mathcal{V}_{k}^{\textrm{F}}}|n,k\rangle|^2}{\hbar \omega_{0} (n-m)}
= -\hbar \frac{4\alpha^2 \Omega_{\mathrm{eff}}^2}{\omega_0 N\nu},
\end{equation}
since $\overline{\mathcal{V}_{k}^{\textrm{F}}}$ only coupled neighboring LLs (dipole selection rule). Let us now consider a truncation of the ladder at LL $n = \Lambda$ (i.e., all levels up to $n=\Lambda$ are included). In this case, the $2p$-th order corrections to the ground-state energy satisfy
\begin{equation}
\mathcal{E}_n^{(2p)} = 0 \quad \text{for} \quad 1 < p \leq \Lambda-n.
\end{equation}
This result can be understood from the structure of perturbation theory. A contribution at order $2p$ corresponds to virtual processes in which the system transitions from the $n$-th LL up to the $(n+p)$-th LL and back. For $p \leq \Lambda - n$, such processes do not probe the cutoff and therefore coincide with those of the infinite ladder, yielding identical (vanishing) contributions. By contrast, when the ladder is truncated at $n = \Lambda - 1$, virtual transitions involving the $\Lambda$-th level are no longer allowed. Consequently, the cancellation mechanism present in the infinite system is broken, and the first non-vanishing correction arises at order $2(\Lambda - n)$. The $n$-th energy level thus acquires a cutoff-dependent contribution of the form
\begin{equation}
\mathcal{E}_n(\alpha) 
= \hbar\omega_0 n 
- \hbar \frac{4\alpha^2 \Omega_{\mathrm{eff}}^2}{\omega_0 N\nu}
+ \frac{K_{2\Lambda-2n}\,\Omega_{\mathrm{eff}}^{2\Lambda-2n}}{N^{\Lambda-n}} \alpha^{2\Lambda-2n}
+ O\!\left(\alpha^{2\Lambda-2n+2}\right),
\end{equation}
where $K_p$ denotes a numerical prefactor arising from the sum of all contributions at order $p$ in the perturbative expansion. Summing over the occupied LLs, $n=0,\dots,\nu-1$, and including the photonic contributions $H_{\textrm{cav}} + H_{\textrm{dia}}$, the total ground-state energy becomes
\begin{align}
E_G(\alpha) 
&= \hbar\left(\omega_0+4D\right)\alpha^2 + N \sum_{n=0}^{\nu-1} \mathcal{E}_n(\alpha) \\
&= -K_2(\Omega_{\mathrm{eff}}^2 - \Omega_c^2)\alpha^2 
+ \frac{K_{2\Lambda-2\nu+2}\,\Omega_{\mathrm{eff}}^{2\Lambda-2\nu+2}}{N^{\Lambda-\nu}} \alpha^{2\Lambda-2\nu+2}
+ O\!\left(\alpha^{2\Lambda-2\nu+4}\right).
\end{align}
Minimizing $E_G(\alpha)$ with respect to $\alpha$ yields the order parameter
\begin{equation}
\frac{\alpha_0^2}{N}
= \left[
\frac{K_2(\Omega_{\mathrm{eff}}^2 - \Omega_c^2)}
{(\Lambda-\nu+1)\, K_{2\Lambda-2\nu+2}\,\Omega_{\mathrm{eff}}^{2\Lambda-2\nu+2}}
\right]^{\frac{1}{\Lambda-\nu}}.
\end{equation}
This result shows that the critical behavior explicitly depends on the cutoff $\Lambda$. Therefore, the resulting critical exponents are not intrinsic to the physical system, but instead arise from the artificial truncation of the Hilbert space. This is a spurious effect; in fact, it has been shown that a rigid cutoff can indeed break gauge invariance~\cite{Chirolli2012}.

\section{Breakdown of the two-level approximation in the superradiant phase}
\label{breakdown_2L}

In this section, we demonstrate that restricting the Hilbert space to the two LLs \( n=\nu-1 \) and \( n=\nu \) is no longer justified in the superradiant phase. We start from the second-quantized electronic Hamiltonian under off-resonant Floquet driving, including the band nonparabolicity, which corresponds to Eq.~(7) of the main text, $\hat{\overline{\mathcal{H}_{k}^{\mathrm{F}}}}=\hat{\mathcal{H}}_{0,k} +\hat{\overline{\mathcal{V}^{\textrm{F}}_k}}$, with 
\begin{align}
\label{single_part_H}
\hat{\mathcal{H}}_{0,k} &= \hbar \omega_0 \sum_n \xi_n\hat{c}_{n,k}^\dagger \hat{c}_{n,k} \nonumber \\
\hat{\overline{\mathcal{V}^{\textrm{F}}_k}} &=  \frac{2 \alpha \hbar \Omega_{\mathrm{eff}}}{\sqrt{N}}  \sum_n f_n\sqrt{\frac{n+1}{\nu}}\left( \hat{c}_{n,k}^\dagger \hat{c}_{n+1,k}+ \hat{c}_{n+1,k}^\dagger \hat{c}_{n,k}\right).
\end{align}
where $\xi_n$ and $f_n$ describe the band nonparabolicity and are defined in the main text. The many-body ground state $|G_{k}\rangle$ can then be expanded to first order in perturbation as
\begin{equation}
|G_{k}\rangle 
= |G^{(0)}_{k}\rangle 
+ \sum_{n} |ne_{k}^{(0)}\rangle 
\frac{\langle ne_{k}^{(0)}| \hat{\overline{\mathcal{V}^{\textrm{F}}_k}} |G^{(0)}_{k}\rangle}{\mathscr{E}_0 - \mathscr{E}_n},
\label{grnd_exp}
\end{equation}
where $\mathscr{E}_0$ and $\mathscr{E}_{n}$ denote the energy of the many-body states $|G^{(0)}_{k}\rangle$ and $|ne_{k}^{(0)}\rangle$, respectively, and do not depend on $k$. The noninteracting ground state is denoted by
\[
|G^{(0)}_{k}\rangle = \prod_{n=0}^{\nu-1} \hat{c}^{\dagger}_{n,k} |0\rangle.
\]
The states $|ne_{k}^{(0)}\rangle$ represent the $n$-th excited states at zeroth order. For instance, the first excited state can be written as $|1e^{(0)}_{k}\rangle = \hat{c}^\dagger_\nu \hat{c}_{\nu-1} |G^{(0)}_{k}\rangle$, while there are two degenerate second excited states: $|2e^{(0)}_{k}\rangle = \hat{c}^\dagger_\nu \hat{c}_{\nu-2} |G^{(0)}_{k}\rangle$ and $|2e^{(0)}_{k}\rangle = \hat{c}^\dagger_{\nu+1} \hat{c}_{\nu-1} |G^{(0)}_{k}\rangle$. Owing to the dipole selection rule imposed by the Hamiltonian, only the first excited state $|1e^{(0)}\rangle$ contributes at first order, which leads to 
\begin{equation}
|G_{k}\rangle 
= |G^{(0)}_{k}\rangle 
- \frac{2 \alpha f_{\nu-1}\Omega_{\mathrm{eff}}}{\omega_0 \sqrt{N}} \, |1e^{(0)}_{k}\rangle.
\end{equation}
From this expression, the occupation probability of the first excited LL $n=\nu$, corresponding to an electron promoted from $n=\nu-1$ to $n=\nu$, is
\begin{equation}
\left| \langle 1e^{(0)}_{k} | G_{k} \rangle \right|^2 
= \left( \frac{2\alpha f_{\nu-1} \Omega_{\mathrm{eff}}}{\omega_0\sqrt{N}} \right)^2.
\end{equation}
Extending the perturbative expansion to higher orders shows that the population of higher excited states scales as
\begin{equation}
\left| \langle ne_{k}^{(0)} | G_{k} \rangle \right|^2
\sim \left( \frac{\alpha \Omega_{\mathrm{eff}}}{\omega_0 \sqrt{N}} \right)^{2n}.
\end{equation}
Therefore, when the photon population is much smaller than the LL degeneracy, i.e., $\alpha^2 \ll N$ as in the normal phase, higher-order excitations are strongly suppressed compared to the first-order contribution. In this regime, the two-level approximation, i.e., restricting the Hilbert space to $n=\nu-1$ and $n=\nu$, remains valid. By contrast, in the superradiant phase the photon population becomes comparable to the LL degeneracy, $\alpha^2 \sim N$, and one typically has $\Omega_{\mathrm{eff}} \sim \omega_0$. As a result, the expansion parameter $\alpha \Omega_{\mathrm{eff}}/(\omega_0 \sqrt{N}) \sim 1$, indicating the breakdown of perturbation theory. The electronic population then spreads nontrivially over many Landau levels, and the full ladder must be taken into account.

\section{Calculation of the expectation values}
\label{numerical_details} 

\begin{figure}
    \centering
    \includegraphics[width=0.4\linewidth]{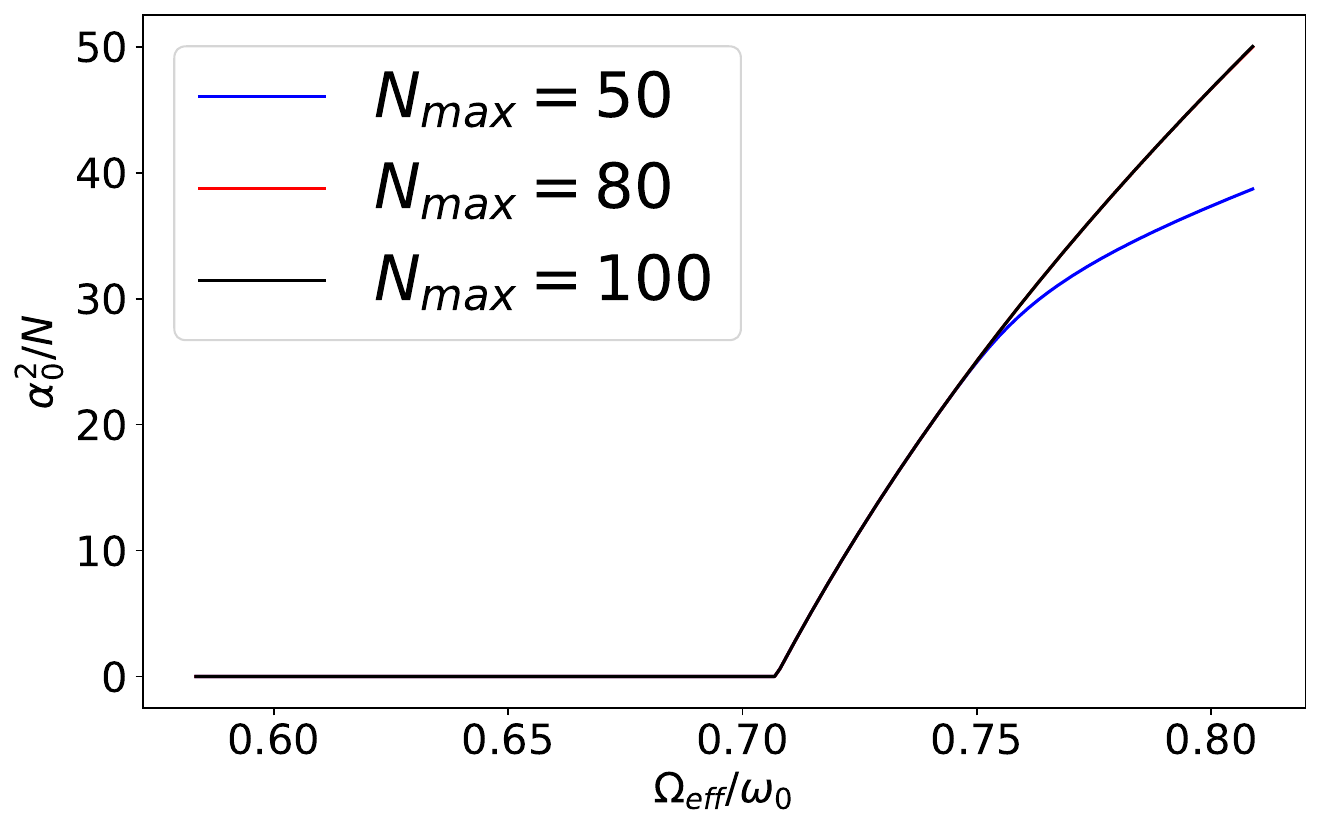}
    \caption{Rescaled photonic order parameter $\alpha_0^2/N$ calculated using energy level truncation $N_{\textrm{max}}=50,80,100$, showing that the results already converge at $N_{\textrm{max}}=80$. Parameters: $\nu=2$ and $\gamma=0.002\omega_0$.}
    \label{fig:alpha0_consistency}
\end{figure}

\begin{figure}
    \centering
    \includegraphics[width=0.4\linewidth]{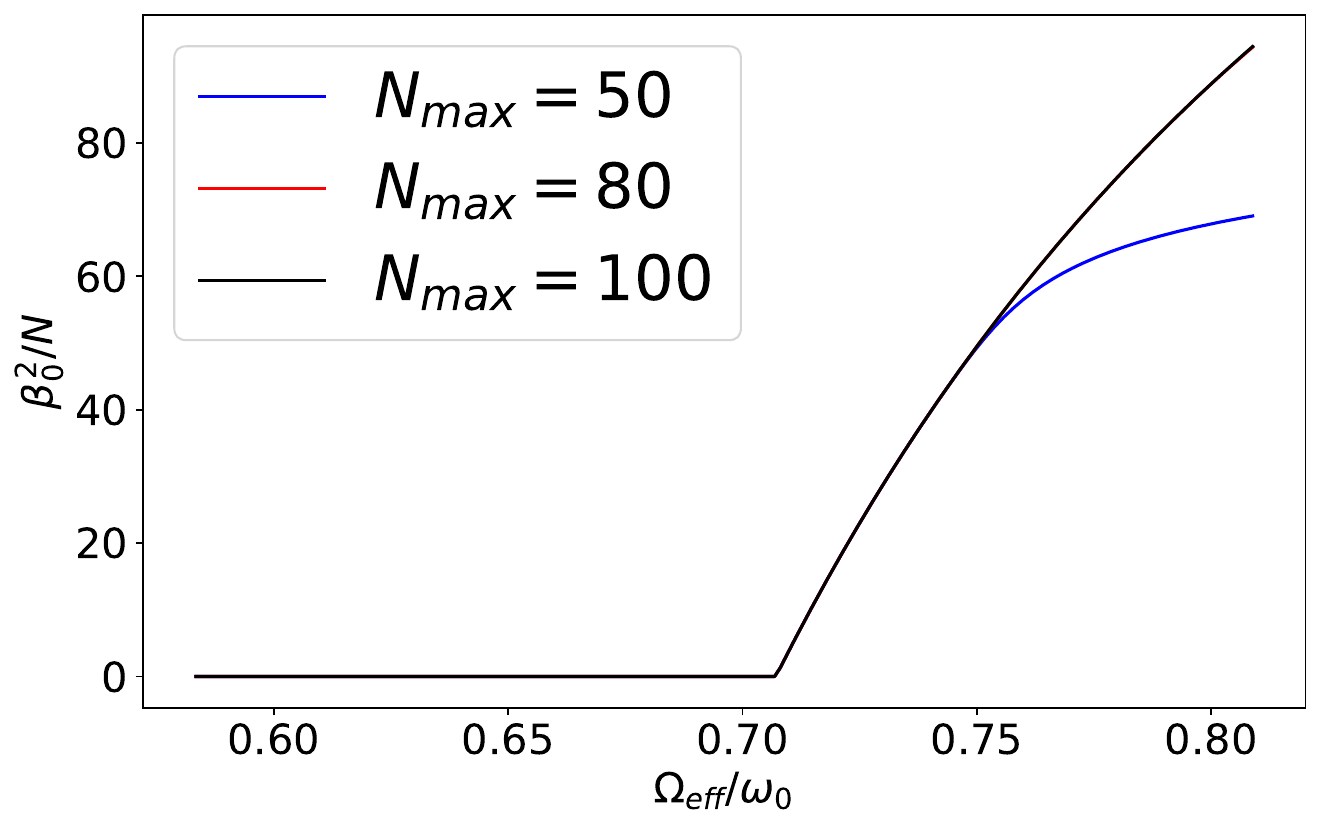}
    \caption{Rescaled excitation number $\beta_0^2/N$ calculated using energy level truncation $N_{\textrm{max}}=50,80,100$, showing that the results already converge at $N_{\textrm{max}}=80$. Parameters: $\nu=2$ and $\gamma=0.002\omega_0$.}
    \label{fig:beta0_consistency}
\end{figure}

In this section, we present the numerical procedure used to compute the order parameters in the presence of band non-parabolicity. The electronic Hamiltonian $\overline{\mathcal{H}^{\textrm{F}}_k}$ given by Eq.~(7) of the main text can be written in a single-electron form as $\overline{\mathcal{H}^{\textrm{F}}_k} = \mathcal{H}_{0,k} + \overline{\mathcal{V}^{\textrm{F}}_k}$, where
\begin{align}
\mathcal{H}_{0,k}  &= \sum_{n=0}^{\infty} \hbar \xi_{n} \, d^{\dagger}_{k} d_{k}, \nonumber \\
\overline{\mathcal{V}^{\textrm{F}}_k} &= \sum_n 2\alpha \hbar \Omega_{\textrm{eff}} \sqrt{\frac{n+1}{\nu N}} f_{n} \left( d_{k} + d^{\dagger}_{k} \right).
\label{eq:numerical_micro}
\end{align}
The eigenvalues $\mathcal{E}_m$ and eigenstates $\{\, |\psi_{m,k}\rangle \,\}$ of $\overline{\mathcal{H}^{\textrm{F}}_k}$ are obtained by diagonalizing Eq.~\eqref{eq:numerical_micro} in the number-state basis $\{\, |n,k\rangle \,\}$. The eigenstates can thus be expressed as linear combinations $|\psi_{m,k}\rangle = \sum_{n} c_{mn} |n,k\rangle$. The Hilbert space is truncated at a finite cutoff $N_{\textrm{max}} =225$, which is sufficiently large to ensure convergence of the numerical results (see Figs.~\ref{fig:alpha0_consistency} and \ref{fig:beta0_consistency}). The many-body ground state is constructed by filling the $\nu$ lowest-energy single-particle eigenstates, $|G_{k}\rangle = |\psi_{0,k}, \psi_{1,k}, \dots, \psi_{\nu-1,k}\rangle$, with corresponding ground-state energy $\mathcal{E}_G = \sum_{m=0}^{\nu-1} \mathcal{E}_m$. We then evaluate $\mathcal{E}_G$ as a function of $\alpha$, and determine the photonic order parameter $\alpha_0$ by minimizing the total ground-state energy $E_G(\alpha) =\hbar(\omega_0 + 4D)\alpha^2 + N \mathcal{E}_G(\alpha)$, i.e., $\alpha_0 = \arg\min_{\alpha} E_G(\alpha)$,
which is computed numerically using Brent's method. Using the resulting many-body ground state $|G_{k}\rangle$, we can compute the electronic polarization, which serves as the matter order parameter and is obtained from the expectation value
\begin{align}
P_{\textrm{2D}}
= -\frac{e}{l_{0}} \sum_{m=0}^{\nu-1} \sum_{n} \frac{\eta\sqrt{2(n+1)\varepsilon}}{\pi} f_n
\langle \psi_{m,k} | 
 d_{k} + d^{\dagger}_{k} 
| \psi_{m,k} \rangle.
\label{matter_op}
\end{align}


\begin{figure}
    \centering
    \begin{minipage}{0.4\textwidth}
        \begin{overpic}[width=\linewidth]{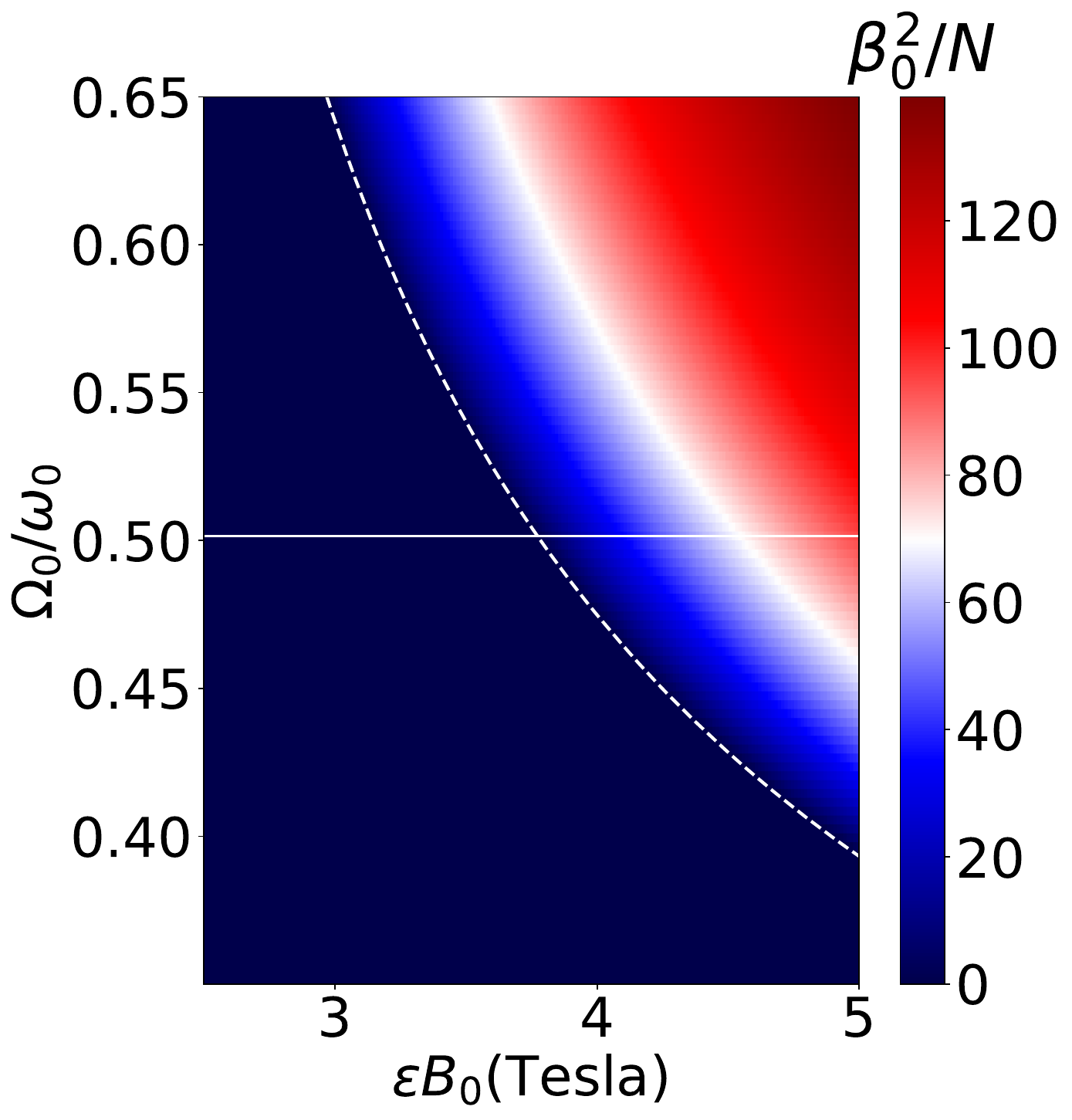}
            \put(20,13){\includegraphics[width=0.32\linewidth]{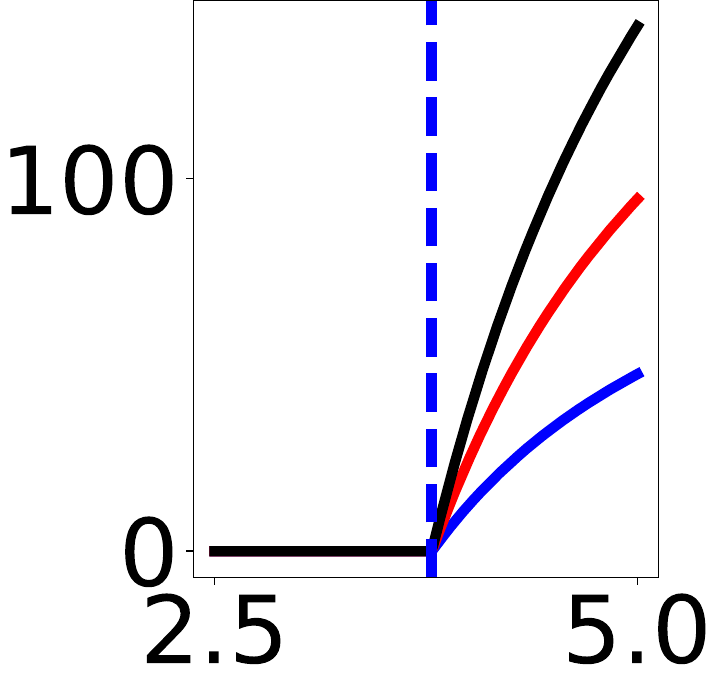}}
        \end{overpic}
    \end{minipage}
    \caption{Rescaled CR excitation number $\beta_0^2/N$ versus $\Omega_{0}/\omega_{0}$ and $\varepsilon B_0$ for $\nu=2$. The phase boundary $\eta^2 \varepsilon_\mathrm{c} = 1+\omega_0^2/(4\Omega_0^2)$ (see main text) is shown as white dashed lines in the figure; insets display horizontal cuts at $\Omega_0/\omega_0 = 0.5$. Blue, red, and black curves correspond to filling factors $\nu = 1, 2,$ and $3$, respectively, while dashed blue lines mark the critical amplitude $\varepsilon = \varepsilon_\mathrm{c}$. }
    \label{fig:beta_2D}
\end{figure}

It is also interesting to compute the excitation number
\begin{equation}
\beta_0^2 = N \sum_{m=0}^{\nu-1} 
\langle \psi_{m,k} | 
\left( \sum_n n d^{\dagger}_{k} d_{k} - n \delta_{mn} \right) | \psi_{m,k} \rangle.
\label{matter_exc_num}
\end{equation}
This quantity is closely related to the magnetization introduced in the Dicke model~\cite{EmaryBrandes2003}. Indeed, within the Holstein--Primakoff representation, the $z$-component of the collective spin of the Dicke model satisfies $J_z=-\frac{N}{2}+b^\dagger b$, where $N$ denotes the number of two-level emitters. The boson number $\langle b^\dagger b\rangle$ therefore measures the deviation of $\langle J_z\rangle$ from its fully polarized ground-state value $-N/2$, $\langle b^\dagger b\rangle = \langle J_z\rangle-\left(-N/2\right)$, and can thus be interpreted as the number of emitters transferred from the ground state to the excited state. The quantity introduced in Eq.~\eqref{matter_exc_num} plays exactly the same role for the LL ladder: $\beta_0^2$ measures the increase of the average LL index relative to the uncoupled ground state. Equivalently, it counts the excess number of cyclotron quanta stored in the electronic system due to the light--matter interaction. Although this quantity remains finite, albeit small, in the normal phase due to ultrastrong coupling~\cite{Ciuti2005}, it becomes of order $N$ in the superradiant phase (see Figure~\ref{fig:beta_2D}).

\section{Approximate expressions of the expectation values}
\label{analytics_order}

In this section, we derive an approximate analytical expression for the photonic order parameter using perturbation theory. We first rewrite the single-electron Hamiltonian in Eq.~\eqref{eq:numerical_micro} in the second-quantized form as $\hat{\overline{\mathcal{H}_{k}^{\mathrm{F}}}}=\hat{\mathcal{H}}_{0,k} +\hat{\overline{\mathcal{V}^{\textrm{F}}_k}}$, where
\begin{align}
\hat{\mathcal{H}}_{0,k} &= \sum_{n} \hbar \xi_{n} \hat{c}_{n,k}^\dagger \hat{c}_{n,k}, \nonumber \\
\hat{\overline{\mathcal{V}^{\textrm{F}}_k}} &= \sum_n 2\alpha \hbar \Omega_{\textrm{eff}} \sqrt{\frac{n+1}{\nu N}}
f_n
\left( \hat{c}_{n+1,k}^\dagger \hat{c}_{n,k} + \hat{c}_{n,k}^\dagger \hat{c}_{n+1,k} \right).
\label{eq:perturbation_micro}
\end{align}
The many-body ground-state energy $\mathcal{E}_G(\alpha) = \langle G_{k} | \hat{\overline{\mathcal{H}_{k}^{\mathrm{F}}}} | G_{k} \rangle$ admits the perturbative expansion 
\begin{align}
\mathcal{E}_G(\alpha) 
&= \mathcal{E}_G^{(0)} + \mathcal{E}_G^{(2)} + \mathcal{E}_G^{(4)} + O(\alpha^6), 
\label{eq:EG}
\end{align} 
where 
\begin{align}
\mathcal{E}_G^{(0)} 
&= \langle G^{(0)}_{k} | \hat{\mathcal{H}}_{0,k} | G^{(0)}_{k} \rangle
= \sum_{n=0}^{\nu-1} \hbar \xi_{n} \nonumber \\
\mathcal{E}_G^{(2)} &= \sum_{m} 
\frac{|\langle G^{(0)}_{k} | \hat{\overline{\mathcal{V}^{\textrm{F}}_k}} | me^{(0)}_{k} \rangle|^2}
{\mathscr{E}_0 - \mathscr{E}_m}
= -\frac{4\Omega_{\mathrm{eff}}^2}{\omega_0 N}\alpha^2 \nonumber \\
\mathcal{E}_G^{(4)} &= \sum_{m,\ell,p}
\frac{
\langle G^{(0)}_{k} | \hat{\overline{\mathcal{V}^{\textrm{F}}_k}} | me^{(0)}_{k} \rangle
\langle me^{(0)}_{k} | \hat{\overline{\mathcal{V}^{\textrm{F}}_k}} | \ell e^{(0)}_{k} \rangle
\langle \ell e^{(0)}_{k} | \hat{\overline{\mathcal{V}^{\textrm{F}}_k}} | pe^{(0)}_{k} \rangle
\langle pe^{(0)}_{k} | \hat{\overline{\mathcal{V}^{\textrm{F}}_k}} | G^{(0)}_{k} \rangle
}{
(\mathscr{E}_0 - \mathscr{E}_m)
(\mathscr{E}_0 - \mathscr{E}_\ell)
(\mathscr{E}_0 - \mathscr{E}_p)
} - \mathcal{E}_G^{(2)} 
\sum_{m}
\frac{|\langle G^{(0)}_{k} | \hat{\overline{\mathcal{V}^{\textrm{F}}_k}} | me^{(0)}_{k} \rangle|^2}
{\left(\mathscr{E}_0 - \mathscr{E}_m\right)^2} \nonumber \\
&= \frac{16 U \Omega_{\mathrm{eff}}^4}{\omega_0^3 N}\alpha^4
\label{eq:EG2}
\end{align}
with the function incorporating band nonparabolicity
\begin{align}
U &= 1 - \frac{\nu-1}{\nu}
\left[\frac{\omega_0 - \gamma(2\nu - 3)}{2\omega_0 - 4\gamma(\nu - 1)}\right]
- \frac{\nu+1}{\nu}
\left[\frac{\omega_0 - \gamma(2\nu + 1)}{2\omega_0 - 4\gamma\nu}\right].
\end{align}
The total ground-state energy is given by $E_G(\alpha) = \hbar(\omega_0 + 4D)\alpha^2 + N \mathcal{E}_G(\alpha)$.
Using Eqs.~\eqref{eq:EG} and \eqref{eq:EG2}, the photonic order parameter is obtained by minimizing the ground-state energy $E_G(\alpha)$, i.e., $\alpha_0 = \arg\min_{\alpha} E_G(\alpha)$. The excitation number $\beta^2_{0}$ is obtained by inserting the perturbative expansion of the ground state $|G_k\rangle$, given in Eq.~\eqref{grnd_exp}, into the definition [Eq.~(8) of the main text]. Altogether, this yields
\begin{align}
\alpha_0^2 
&= \frac{\omega_0^2}{8\Omega_{\mathrm{eff}}^4}
\frac{\Omega_{\mathrm{eff}}^2 - \Omega_c^2}{U}, \\
\beta_0^2 
&= \frac{\Omega_{\mathrm{eff}}^2 - \Omega_c^2}{2U \Omega_{\mathrm{eff}}^2},
\end{align}
for $\Omega_{\mathrm{eff}} > \Omega_c$, while $(\alpha_0^2,\beta_0^2) = (0,0)$ for $\Omega_{\mathrm{eff}} < \Omega_c$.

\bibliography{references}